\documentclass[reprint, aps, pra, footinbib, letterpaper, superscriptaddress]{revtex4-1}

\usepackage[T1]{fontenc} 

\usepackage{amsmath,color}
\usepackage{amssymb}
\usepackage{amsfonts}
\usepackage{amsthm}
\usepackage{mathtools}
\usepackage{leftidx}

\usepackage[title]{appendix}

\usepackage{bbold}
\usepackage{algorithm}
\usepackage{algpseudocode}
\usepackage{dsfont}

\DeclarePairedDelimiter\bra{\langle}{\rvert}
\DeclarePairedDelimiter\ket{\lvert}{\rangle}
\DeclarePairedDelimiterX\braket[2]{\langle}{\rangle}{#1 \delimsize\vert #2}

\newcommand{\vJ}{\mathbf{J}}
\newcommand{\vP}{{\boldsymbol{\mathcal{P}}}}

\newcommand{\vE}{\mathbf{E}}
\newcommand{\vEin}{\mathbf{E}_{\text{in}}}

\newcommand{\vB}{\mathbf{B}}
\newcommand{\vGreen}{\overleftrightarrow{\mathbf{G}}}
\newcommand{\veta}{\overleftrightarrow{\mathbf{\eta}}}
\newcommand{\vI}{\overleftrightarrow{\mathbf{I}}}
\newcommand{\vD}{\mathbf{D}}
\newcommand{\ve}{\mathbf{e}}
\newcommand{\vnabla}{\boldsymbol{\nabla}}
\newcommand{\vdelta}{\boldsymbol{\delta}}
\newcommand{\vmu}{\boldsymbol{\mu}}
\newcommand{\hmu}{\hat{\boldsymbol{\mu}}}

\newcommand{\vR}{\mathbf{r}}
\newcommand{\vRR}{\mathbf{R}}

\newcommand{\kFGR}{k_{\text{FGR}}}

\newcommand{\kEh}{k_{\text{Eh}}}

\newcommand{\kET}{k_{\text{ET}}}

\newcommand{\he}{\mathbf{e}}
\newcommand{\hH}{\hat{H}}
\newcommand{\hQ}{\hat{\mathcal{Q}}}

\newcommand{\hV}{\hat{V}}
\newcommand{\hsigma}{\hat{\sigma}}

\newcommand{\hP}{\hat{\boldsymbol{\mathcal{P}}}}

\newcommand{\hrho}{\hat{\rho}}

\newcommand{\Us}{U_{\text{s}}}

\newcommand{\tr}[1]{\text{Tr}\left(#1\right)}

\newcommand{\Lrhosub}[2]{\mathcal{L}_{\text{#1}}[\hat{\rho}_\text{#2}]}

\newcommand{\avg}[1]{\left\langle #1\right\rangle}

\newcommand{\real}[1]{{\text{Re}\left[#1\right] }}
\newcommand{\imag}[1]{{\text{Im}\left[#1\right] }}

\begin{document}

	\title{Understanding the Nature of Mean-Field Semiclassical Light-Matter Dynamics: An Investigation of Energy Transfer, Electron-Electron Correlations, External Driving and Long-Time Detailed Balance}%
	
	\author{Tao E. Li}%
	\email{taoli@sas.upenn.edu}
	\affiliation{Department of Chemistry, University of Pennsylvania, Philadelphia, Pennsylvania 19104, USA}

	\author{Hsing-Ta Chen}
	\affiliation{Department of Chemistry, University of Pennsylvania, Philadelphia, Pennsylvania 19104, USA}
	
	\author{Abraham Nitzan} 
	\affiliation{Department of Chemistry, University of Pennsylvania, Philadelphia, Pennsylvania 19104, USA}

	\author{Joseph E. Subotnik}
	\email{subotnik@sas.upenn.edu}
	\affiliation{Department of Chemistry, University of Pennsylvania, Philadelphia, Pennsylvania 19104, USA}

	\begin{abstract}
		Semiclassical electrodynamics [with quantum matter plus classical electrodynamics fields] is an appealing approach for studying light-matter interactions, especially for realistic molecular systems. However, there is no unique semiclassical scheme. On the one hand, intermolecular interactions can be described instantaneously by static two-body interactions connecting two different molecules, while a classical transverse E-field acts as a spectator at short distance; we will call this Hamiltonian \#I. On the other hand, intermolecular interactions can also be described as effects that are mediated exclusively through a classical one-body E-field without any quantum effects at all (assuming we ignore electronic exchange); we will call this Hamiltonian \#II. Moreover, one can also mix these two different Hamiltonians into a third, hybrid Hamiltonian, which preserves quantum electron-electron correlations for lower excitations but describes higher excitations in a mean-field way. To investigate which semiclassical scheme is most reliable for practical use, here we study the real-time dynamics of a minimalistic many-site model --- a pair of identical two-level systems (TLSs) --- undergoing either resonance energy transfer (RET) or collectively driven dynamics. While both approaches perform reasonably well (\#1 as \#2) when there is no strong external excitation, we find that no single approach is perfect for all conditions (and all methods fail when a strong external field is applied). Each method has its own distinct problems: Hamiltonian \#I performs best for RET but behaves in a complicated manner for driven dynamics. Hamiltonian \#II is always stable, but obviously fails for RET at short distances. One key finding is that, for externally driven dynamics, a full configuration interaction description of Hamiltonian \#I (\#I-FCI) strongly overestimates the long-time electronic energy, highlighting the not obvious fact that, if one plans to merge quantum molecules with classical light, a full, exact treatment of electron-electron correlations can actually lead to worse results than a simple mean-field electronic structure treatment. Future work will need to investigate (i) how these algorithms behave in the context of more than a pair of TLSs and (ii) whether or not these algorithms can be improved in general by including crucial aspects of spontaneous emission.
	\end{abstract}

	\maketitle

	\section{Introduction}\label{sec:intro}
	
	Recent experiments demonstrating collective phenomena with nanoscale light-matter interactions\cite{Torma2015,Jia2018,Schafer2019} have highlighted the need for computational simulations of 
	realistic molecular systems\cite{Thakkar2015,Du2018,Flick2017}.
	Unfortunately, full quantum electrodynamical (QED) calculations scale unfavorably with the number of quantized photonic modes. Moreover, full QED is compatible only with  full configuration interactions (CI) for the description of the matter system, such that QED also scales unfavorably with the number of molecules.
	Thus, mixed quantum--classical electrodynamics are a promising approach with reduced computational cost: one treats electronic/molecular subsystems with approximate quantum mechanics and describes light fully classically. For decades, semiclassical electrodynamical simulations have captured many exciting phenomena in the field of quantum optics and spectroscopy\cite{Gross1982,Puthumpally-Joseph2014,Puthumpally-Joseph2015,Sukharev2011,Neuhauser2007,Lopata2009-1,Lopata2009-2,Lisinetskaya2014}.
	
	Nevertheless, semiclassical electrodynamics suffers from many well-known issues. First, vacuum fluctuations are ignored due to the classical treatment of EM fields, which are usually calculated via a mean-field (Ehrenfest) approximation. Owing to the failure of the classical EM field description, semiclassical electrodynamics cannot fully recover any pure quantum effects for a single electronic system, such as spontaneous emission\cite{Crisp1969,Milonni1976}. To date,  many techniques have been proposed to account for this issue and this work is still continuing\cite{Flick2017,Li2018Spontaneous,Chen2018Spontaneous,Hoffmann2019,Hoffmann2018}.
	
	The second issue for semiclassical electrodynamics is that there is no unique semiclassical Hamiltonian, and inevitably, some inconsistency must arise because of the semiclassical ansatz.
	After all, how should we treat electron-electron interactions? Are they instantaneous and static? Are they mediated exclusively by the EM field or not? If one chooses a static picture, one assumes the electronic Hamiltonian is a combination of quantum two-body terms plus an electric dipole coupling term (which defines Hamiltonian \#I in Ref. \citenum{Li2018Tradeoff}); here one finds that 
	one can predict accurate short-range RET rate but at the cost of violating the  long-range causality due to a quantum-classical mismatch of intermolecular interactions\cite{Li2018Tradeoff}.
	By contrast, if one chooses to couple matter exclusively through the field, one assumes the electronic Hamiltonian will have only an extended dipole coupling term  (which defines Hamiltonian \#II); here one finds that one fails to capture any short-range RET rate quantitatively due to the lack of quantum electron-electron correlations but strictly preserves causality\cite{Li2018Tradeoff}.  
	
	The problem of correlation versus causality  is usually ignored in the literature. 
	Nowadays, almost all calculations use Hamiltonian \#II at the cost of inaccurate short-range interactions\cite{Ziolkowski1995,Sukharev2011,DiLoreto2018,Jestadt2018}.
	Of course, one means of improving Hamiltonian \#II is to use density functional theory (DFT) for electronic structure. In principle, DFT/TD-DFT can give exact electronic structure while maintaining the single-body nature of the electronic Hamiltonian. Beyond DFT, however, there is very little work using explicitly correlated electronic wavefunctions interacting with both external and internal EM fields. In general, if one wants to use explicitly correlated electronic wavefunctions (to account for electron-electron correlations) while studying light-matter interactions, to date the usual premise has been to first diagonalize a molecular electronic Hamiltonian (with no explicit electric field but rather only with
	instantaneous Coulomb terms) and then allow the resulting many-body electronic states to interact with an external electric field\cite{Klamroth2003,Krause2005,Greenman2010,Tremblay2008,Rohringer2006}. As such, the electronic dynamics as induced by internally generated, dynamic electric fields is not usually accounted for. 
	As a result, almost all standard approaches fail to capture some key effects of collective phenomena, for example modification of the spontaneous decay rate, the effect of the dielectric constant, or even the presence of an RET rate\cite{Scheel1999,Lo2001,Acob2018}. 
	For our purposes, we will not invoke DFT in the present paper, and our goal is to establish a clean benchmark of Hamiltonian \#I and \#II dynamics, and distinguish between purely mean-field electronic dynamics and explicitly correlated electronic dynamics. We will attempt to answer the following equations:
	\begin{enumerate}
		\item[(i)] By including quantum electron-electron correlations, is Hamiltonian \#I always  superior to Hamiltonian \#II in practice?
		\item[(ii)] Can we always improve semiclassical results for Hamiltonian \#I by treating quantum electron-electron correlations at a higher level of accuracy?
		For example, in the context of a Hartree-Fork (HF) ground state and configuration interaction singles (CIS) excited states, does the performance always  improve if we increase the size of our configuration interaction (CI) Hamiltonian to include higher excited CIs (e.g., doubly excited CIs)? 
	\end{enumerate}
	In order to answer these questions,  we will investigate resonant energy transfer (RET) and collectively driven electronic dynamics for a minimalistic two-site model  within the framework of mean-field Ehrenfest dynamics. While we have previously applied the same model to study the short-time RET rate\cite{Li2018Tradeoff}, we will  now study long-time RET dynamics as well as the crucial effects of including an external driving field (using a standard dyadic Green's function technique; see Appendix).
	Understanding this minimal model should pave the way for improving the currently available semiclassical methods. 
	
	This paper is organized as follows. In Sec. \ref{sec:method}, we introduce the framework of mean-field Ehrenfest dynamics as well as different semiclassical Hamiltonians. In Sec. \ref{sec:model}, we introduce the model and parameters for simulations. In Sec. \ref{sec:result}, we present results for RET and driven dynamics, showing that some unexpected, anomalous behavior can emerge.
	In Sec. \ref{sec:discussion},   we explain the reasons for this reported anomaly.
	We conclude in Sec. \ref{sec:conclusion}.
	
	
	\section{Method: Semiclassical Electrodynamics}\label{sec:method}
	
	As a brief review, we will now review the conventional semiclassical method --- mean-field (Ehrenfest) dynamics --- for propagating light-matter electrodynamics. First, according to which the matter side obeys the time-dependent Schr\"odinger equation, 
	\begin{equation}
		\label{eq:WaveFunction-Ehrenfest}
		\frac{d}{dt} \ket{\Psi_N(t)} = -\frac{i}{\hbar} \hH_{\text{sc}} \ket{\Psi_N(t)}.
	\end{equation}
	Here, $\ket{\Psi_N}$ denotes the electronic wave function for $N$ molecules, and  $\hH_{\text{sc}}$ denotes the semiclassical Hamiltonian, which will be introduced later. 
	Second, for the EM side, the classical Maxwell's equations are evolved:
	\begin{subequations}
		\label{eq:Maxwell}
		\begin{align}
		\frac{\partial}{\partial t} \vB(\vR, t) &= - \vnabla \times \vE(\vR, t) \\
		\frac{\partial}{\partial t} \vE(\vR, t) &= c^2 \vnabla \times \vB(\vR, t) - \frac{\vJ(\vR, t)}{\epsilon_0},
		\end{align}
	\end{subequations}
	where $\epsilon_0$ denotes the vacuum permittivity.
	Here, the current density $\vJ(\vR, t)$ is calculated by a mean-field approximation:
	\begin{equation}
		\label{eq:J-Ehrenfest}
		\begin{aligned}
			\vJ(\vR, t) = \sum\limits_{n=1}^{N}\frac{\partial }{\partial t}\tr{\hrho(t)\hP^{(n)}(\vR)}
		\end{aligned}
	\end{equation}
	Here, $\hP^{(n)}$ denotes the polarization density operator for molecule $n$.
	Eqs. \eqref{eq:WaveFunction-Ehrenfest}-\eqref{eq:J-Ehrenfest} are called the coupled Maxwell-Schr\"odinger equations.	
	 In this framework, the only remaining question is how to define the form of $\hH_{\text{sc}}$.
	
	\subsection{Hamiltonian \#I}
	For neutral and non-overlapping molecules that interact with the E-field solely, the standard semiclassical Hamiltonian  reads\cite{Mukamel1999}
	\begin{equation}
		\label{eq:H-I}
			\hH_{sc}^{I} = \sum_{n = 1}^{N}\hH_s^{(n)} -\int d\vR \ \vE_{\perp}(\vR,t)\cdot \hP^{(n)}(\vR)  
							+ \sum_{n < l} \hV_{\text{Coul}}^{(nl)} 
	\end{equation}
	Here, $\hH_s^{(n)}$ denotes the molecular Hamiltonian for molecule $n$; molecules interact with each other through a classical transverse E-field $\vE_{\perp}$, and electron-electron correlations between molecules are characterized by the intermolecular Coulomb operator
	\begin{equation}
	\label{eq:vdd_QED}
		\hV_{\text{Coul}}^{(nl)} = \frac{1}{\epsilon_0} \int d\vR \hP^{(n)}_{\parallel}(\vR)\cdot \hP^{(l)}_{\parallel}(\vR)
	\end{equation}	
	We note that $\hV_{\text{Coul}}^{(nl)}$ scales as $1/R^3$ (where $R$ denotes intermolecular separations), $\int d\vR \ \vE_{\perp}(\vR,t)\cdot \hP^{(n)}(\vR)$ scales as $1/R$. Thus, $\hV_{\text{Coul}}^{(nl)}$ dominates short-range intermolecular interactions, while $\vE_{\perp}(\vR,t)$ dominates long-range intermolecular interactions. For usual F\"orster resonance energy transfer (FRET)\cite{Forster1948}, we usually account only for $\hV_{\text{Coul}}^{(nl)}$, leading to a $1/R^6$ dependence of the energy transfer rate (which follows from a Fermi's golden rule calculation).

	According to Eqs. \eqref{eq:H-I} and \eqref{eq:vdd_QED}, the exchange operator between molecules is neglected, which is adequate when the wave functions between molecules do not overlap. In this manuscript, we will call Eq. \eqref{eq:H-I}  Hamiltonian \#I. In general, for $N$ TLSs, the quantum two-body term $\hV_{\text{Coul}}^{(nl)}$ introduces a great deal of the computational complexity. Hamiltonian \#I formally should require a Hilbert space of size $2^N$. Thus, in practice, when modeling electrodynamics, one is forced to construct approximations to Hamiltonian \#I of which there are many. We will now investigate two such variants with different electronic structure theories to propagate the time-dependent Schr\"odinger equation.
	

	\subsubsection{Time-Dependent Full Configuration Interaction}
	To fully account for $\hV_{\text{Coul}}^{(nl)}$, if one has the means, one can propagate the time-dependent Schr\"odinger equation in a complete basis using Hamiltonian \#I. These exact, molecular quantum dynamics are known as  time-dependent full configuration interaction (TD-FCI). Obviously,  TD-FCI is  possible only for simple models, i.e., a few two-level systems (TLSs), as $2^N$ grows fast for large $N$.
	
	\subsubsection{Time-Dependent Configuration Interaction Singles}
	For large systems with many molecules, in order to reduce the computational cost, the most common treatment is to truncate Hamiltonian \#I at the level of  single excitations, also called the time-dependent configuration interaction singles (TD-CIS) method. Here, the time-dependent electronic wave function is expanded as:
	\begin{subequations}
		\label{eq:Psi_CIS_tot}
		\begin{equation}
		\label{eq:Psi_CIS}
		\ket{\Psi_N(t)} \approx \ket{\Psi_{\text{CIS}}(t)} = \sum_{I}C_I(t)\ket{\Psi_I}
		\end{equation}
	where $C_I(t)$ is a time-dependent coefficient and $\ket{\Psi_I}$ denotes the $I$-the CIS state, which is defined as
	\begin{equation}
	   \ket{\Psi_I} = D_{0,I}\ket{\Psi_0^{\text{HF}}} + \sum_{i = L}^{\frac{N_e}{2}}\sum_{a=\frac{N_e}{2}+1}^{M}D_{I,i}^a \ket{\Psi_i^{a}}
	\end{equation}
	Here, $\ket{\Psi_0^{\text{HF}}}$ denotes the restricted Hartree-Fock ground state for the electronic degrees of freedom in the absence of EM fields, $\ket{\Psi_i^{a}}$ denotes a singly excited state by exciting an electron from an occupied molecular orbital (MO) $i$ to an unoccupied MO $a$, $N_e$ denotes the number of electrons, $L$ is the lowest occupied orbital, and $M$ is the highest unoccupied orbital.
	\end{subequations}
	
	Given the CIS wavefunction that is defined in Eq. \eqref{eq:Psi_CIS_tot}, one can propagate the wavefunction as
	\begin{equation}\label{eq:CIS_EOM}
	\frac{d}{dt}C_I(t) = -\frac{i}{\hbar} \sum_{J} \avg{\Psi_I\bigg |\hH_{\text{sc}}^{I} \bigg | \Psi_J}C_J(t)
	\end{equation}
	where $\hH_{\text{sc}}^{I}$ is already defined in Eq. \eqref{eq:H-I}.

	\subsection{Hamiltonian \#II}
	Even simpler than TD-CIS, a more radical solution is to invoke the mean-field approximation (or Hartree approximation) for $\hV_{\text{Coul}}^{(nl)}$ [Eq. \eqref{eq:vdd_QED}]:
	\begin{widetext}
	\begin{subequations}
	\begin{align}
		\label{eq:Hartree-approximation-vdd-pre}
		\hV_{\text{Coul}}^{(nl)} &\approx \frac{1}{\epsilon_0} \int d\vR \left[  \vP^{(n)}_{\parallel}(\vR, t)\cdot \hP^{(l)}_{\parallel}(\vR) + \vP^{(l)}_{\parallel}(\vR, t)\cdot \hP^{(n)}_{\parallel}(\vR)\right] - \frac{1}{\epsilon_0} \int d\vR \vP^{(n)}_{\parallel}(\vR, t)\cdot \vP^{(l)}_{\parallel}(\vR) \\
		\label{eq:Hartree-approximation-vdd}
		&\approx  \frac{1}{\epsilon_0} \int d\vR \left[  \vP^{(n)}_{\parallel}(\vR, t)\cdot \hP^{(l)}_{\parallel}(\vR) + \vP^{(l)}_{\parallel}(\vR, t)\cdot \hP^{(n)}_{\parallel}(\vR)\right] \ \ \ \text{(up to a constant)}   
	\end{align}
	\end{subequations}
	\end{widetext}
	where $\vP^{(n)}_{\parallel}(\vR, t)$ denotes the longitudinal component of the classical polarization density for molecule $n$. 
	Keen readers might well be confused about the mean-field treatment in Eq. \eqref{eq:Hartree-approximation-vdd-pre}: Why not take  $\hV_{\text{Coul}}^{(nl)} \approx \frac{1}{2\epsilon_0} \int d\vR \left[  \vP^{(n)}_{\parallel}(\vR, t)\cdot \hP^{(l)}_{\parallel}(\vR) + \vP^{(l)}_{\parallel}(\vR, t)\cdot \hP^{(n)}_{\parallel}(\vR)\right]$ instead? The motivation behind Eq. \eqref{eq:Hartree-approximation-vdd-pre} is twofold: (i) Eq. \eqref{eq:Hartree-approximation-vdd-pre}  allows us to define a semiclassical Hamiltonian that strictly preserves causality, as is shown below; (ii) the mean-field expansion in Eq. \eqref{eq:Hartree-approximation-vdd-pre} is already standard  in the area of many-body physics; see Ref. \citenum{bruus2002introduction} for a brief introduction. Because the last term in Eq. \eqref{eq:Hartree-approximation-vdd-pre} is just a time-dependent constant and will not alter the equations of motion for the molecular part, this term can be further neglected, leading to Eq. \eqref{eq:Hartree-approximation-vdd}.
	
	By substituting Eq. \eqref{eq:Hartree-approximation-vdd} into Eq. \eqref{eq:H-I}, we arrive at Hamiltonian \#II:
	\begin{equation}
	\label{eq:H-II}
	\begin{aligned}
	\hH_{sc}^{II} =  &\sum_{n = 1}^{N} \hH^{(n)}_\text{MF}.
	\end{aligned}
	\end{equation}
	where 
	\begin{equation}
	\label{eq:H-MF}
	\begin{aligned}
	\hH^{(n)}_\text{MF} =  & \hH_s^{(n)} -\int d\vR \ \vE(\vR,t)\cdot \hP^{(n)}(\vR) \\
	&+ \frac{1}{\epsilon_0} \int d\vR \vP^{(n)}_{\parallel}(\vR, t)\cdot \hP^{(n)}_{\parallel}(\vR)
	\end{aligned}
	\end{equation}
	Within  Hamiltonian \#II,  molecules interact with each other only through a classical E-field $\vE(\vR,t)$, and the last term above [in Eq. \eqref{eq:H-MF}] denotes the semiclassical self-polarization, which effectively renormalizes the energy levels of molecules slightly and does not significantly alter the overall dynamics. Hence, we will neglect the last term  in our numerical simulations. Keen readers might wonder  whether energy conservation is still valid  if the last term is neglected --- indeed, energy conservation can be guaranteed if we simply redefine the conserved quantity\cite{Li2018Tradeoff}.
	
	
	\subsubsection{Time-dependent Hartree Method}
	Given the one-body nature of Hamiltonian \#II, the time-dependent Schr\"odinger equation can be evolved exactly with simple time-dependent Hartree (TDH) dynamics, i.e., the electronic wavefunction can be written as a Hartree product:
	\begin{equation}
	\label{eq:Psi_Hartree}
	    \ket{\Psi_N(t)} = \ket{\psi_{1}(t)}\ket{\psi_{2}(t)}\cdots\ket{\psi_{N}(t)}
	\end{equation}
	where $\ket{\psi_{n}(t)}$ denotes an effective one-body wavefunction for molecule $n = 1, 2, \cdots, N$. Following the variational principle\cite{Beck2000}, the equation of motion for each orbital $\ket{\psi_{n}(t)}$ can be obtained as:
	\begin{equation}
		\label{eq:EOM-TDH}
		\frac{d}{dt} \ket{\psi_{n}(t)} = -\frac{i}{\hbar} \hH^{(n)}_\text{MF} \ket{\psi_{n}(t)}
	\end{equation} 
	where $\hH^{(n)}_\text{MF}$ is defined in Eq. \eqref{eq:H-MF}, and $\vP^{(n)}_{\parallel}(\vR, t) = \bra{\psi_{n}(t)}\hP_{\parallel}^{(n)}(\vR)\ket{\psi_{n}(t)}$.

	\subsection{A Hybrid Hamiltonian}
	While Hamiltonian \#I treats only the transverse E-field classically, Hamiltonian \#II treats all intermolecular interactions classically. Interestingly, we can write both of these Hamiltonians in  a uniform way:
	\begin{equation}
	\label{eq:H-hybrid}
	\begin{aligned}
	\hH_{sc} =  \hH_{sc}^{II} + \hQ\delta\hV_{\text{Coul}}\hQ
	\end{aligned}
	\end{equation}
	where  $\delta\hV_{\text{Coul}}$ is defined as
	\begin{equation}
		\label{eq:delta_VCoul}
	\begin{aligned}
		\delta\hV_{\text{Coul}} \equiv &\sum_{n\neq l} \hV_{\text{Coul}}^{(nl)} - \frac{1}{\epsilon_0} \int d\vR \large[  \vP^{(n)}_{\parallel}(\vR, t)\cdot \hP^{(l)}_{\parallel}(\vR) \\
		&+ \vP^{(l)}_{\parallel}(\vR, t)\cdot \hP^{(n)}_{\parallel}(\vR)\large]
	\end{aligned}
	\end{equation}
	and $\hQ$ denotes a projection operator into a subspace ($W$) of the electronic states:
	\begin{equation}
	\label{eq:hQ}
	\hQ \equiv \sum\limits_{i\in W}\ket{i}\bra{i}
	\end{equation}
	When $W = \{\emptyset \}$, $\hQ = \mathbb{0}$, and 
	Eq. \eqref{eq:H-hybrid} reduces to Hamiltonian \#II; when $W$ is the entire electronic manifold of states $\mathcal{S}$, $Q = \mathds{1}$, and Eq. \eqref{eq:H-hybrid}  reduces to Hamiltonian \#I. By choosing an arbitrary subspace in between $\{\emptyset \}$ and $\mathcal{S}$, we can find intermediate Hamiltonians in between Hamiltonians \#I and \#II.
	Hence, Eqs. \eqref{eq:H-hybrid}-\eqref{eq:hQ} form a generalized definition of a semiclassical Hamiltonian. Clearly, the choice of the subspace will play an  important role in the quality of the Hamiltonian.
	In this manuscript, we define one intermediate subspace as:
	\begin{equation}
	\label{eq:W_subspace}
	 W_{0+1}  = \{ \text{ground or single excitonic states} \}
	\end{equation}
	We call Eqs. \eqref{eq:H-hybrid}-\eqref{eq:W_subspace} a hybrid Hamiltonian ($\hH_{\text{sc}}^{hyb}$), in which there are two-body   couplings ($\hV_{\text{Coul}}^{(nl)}$) for the ground and singly excited states, but the double and higher excited states are entirely decoupled and reduced to mean-field interactions. 
	
	\subsubsection{Time-dependent Hybrid Method}
	For the hybrid Hamiltonian, the many-body wave function can be expanded as follows:
	\begin{equation}
	\ket{\Psi_N(t)} = \ket{\Psi_{\text{CIS}}(t)}\otimes\ket{\psi_{he}(t)}
	\end{equation}
	Here, $\ket{\Psi_{\text{CIS}}(t)}$ characterizes the wave function for the CIS states, which is defined in Eq. \eqref{eq:Psi_CIS_tot}, and $\ket{\psi_{he}(t)}$ characterizes the wave function for higher excitations. On the one hand, we evolve $\ket{\Psi_{\text{CIS}}(t)}$ by TD-CIS as in Eq. \eqref{eq:CIS_EOM}; on the other hand, because each higher excited state interacts with other states (i.e., CIS and other higher excited states) solely through a classical E-field, these states can be propagated independently with TDH as in Eq. \eqref{eq:EOM-TDH}. For example, for a pair of TLSs, the explicit form of the hybrid Hamiltonian is presented in Eq. \eqref{eq:Hhyb_2TLS}.
	
	
	\subsection{Summary of Semiclassical Hamiltonians}
	
	\begin{figure}
		\includegraphics[width=1.0\linewidth]{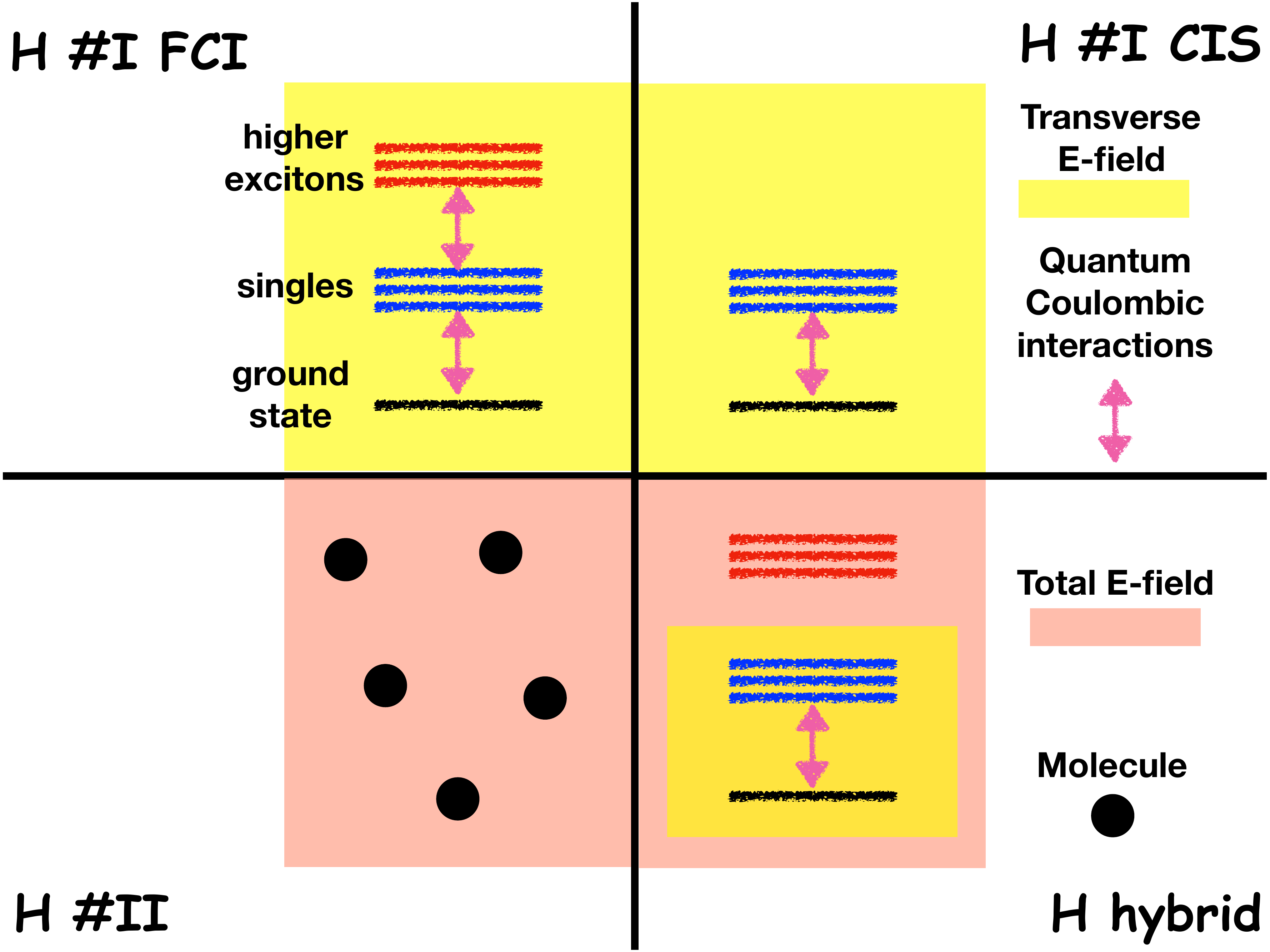}
		\caption{Cartoon of four semiclassical approaches: Hamiltonian \#I FCI, \#I CIS, \#II, and the hybrid Hamiltonian. Intermolecular interactions are incorporated by (i) quantum intermolecular Coulomb interactions plus a classical transverse E-field for Hamiltonian \#I; (ii) a classical total E-field for Hamiltonian \#II.   (iii) For the hybrid Hamiltonian, the ground state and singles are treated with Hamiltonian \#I CIS, while higher excitations interact with others (and themselves) through a classical E-field solely.}
		\label{fig:four_methods}
	\end{figure}

\begin{table*}
	\caption{Synopsis of the main features of the semiclassical Hamiltonians  for modeling light-matter interactions}
	\label{table:Hamiltonians}
	\centering
	\begin{tabular}{lcccr}
		\hline
		Approach & Definition & Quantum e-e correlations \ \ \ & Computational complexity \ \ \ & Causality
		\\ \hline 
		\#I FCI & Eqs. \eqref{eq:H-I}-\eqref{eq:vdd_QED} & Fully accounted  &   $O(2^{N})$& Violated
		\\ 
		\#I CIS & Eqs. \eqref{eq:H-I}-\eqref{eq:Psi_CIS}  & Partially accounted& $O(N^2)$ &Violated \\ 
		\#II &  Eqs. \eqref{eq:H-II}-\eqref{eq:H-MF}  & None& $O(N)$ & Preserved\\ 
		hybrid & Eqs. \eqref{eq:H-hybrid}-\eqref{eq:W_subspace} & Partially accounted &  $O(N^2)$ &Violated
		\\ \hline
	\end{tabular}
	\normalsize
\end{table*}

	Fig. \ref{fig:four_methods} is a cartoon of the four different semiclassical approaches (Hamiltonian \#I FCI, \#I CIS, \#II, and a hybrid Hamiltonian) that have been introduced above. In this cartoon, we highlight how intermolecular interactions are described differently in these approaches. 
	
	Table \ref{table:Hamiltonians} also summarizes the important features of these Hamiltonians, e.g., defining equations, whether or not quantum electron-electron correlations are accounted for, computational complexity as a function of molecular number ($N$), and whether long-range causality is preserved or not.

	\section{Model}\label{sec:model}
	From now, natural units will be  used: $[\hbar]=[c]=[\epsilon_0]=1$.
	We will perform calculations with a minimalistic quantum model --- a pair of identical TLSs (labeled as D and A). The molecular Hamiltonian for molecule $n=D,A$ reads:
	\begin{equation}
		\label{eq:Hs}
		\hH_s^{(n)} = \hbar\omega_0\hsigma_{+}^{(n)}\hsigma_{-}^{(n)}
	\end{equation}
	where $\hbar\omega_0$ denotes the energy gap between ground state $\ket{ng}$ and excited state $\ket{ne}$ for molecule $n$, $\hsigma_{+}^{(n)} \equiv \ket{ne}\bra{ng}$, and $\hsigma_{-}^{(n)} \equiv \ket{ng}\bra{ne}$. After the long-wavelength approximation, $\hP^{(n)}(\vR)$ reads:
	\begin{equation}
		\label{eq:polarization-density-operator-molecule}
		\hP^{(n)}(\vR) =  \mu_{ge}\ve_d^{(n)}\delta(\vR-\vR_n)\hsigma_x^{(n)}
	\end{equation}
	where  $\hsigma_x^{(n)} = \ket{ng}\bra{ne} + \ket{ne}\bra{ng}$, $\mu_{ge}$ denotes the magnitude of the transition dipole moment, and $\ve_d^{(n)}$ and $\vR_n$ denote the unit vector along the transition dipole and the position of molecule $n$.

	For our simulation parameters, we suppose that the TLSs are positioned symmetrically at $\vR_n = (0, \pm \frac{R}{2}, 0)$, and their transition dipole moments are both oriented along the $z$-axis ($\ve_d^{(n)} = \ve_z$). We set $\omega_0 = 1$ and $\mu_{ge} = 0.1$. In vacuum, the spontaneous emission rate for a single TLS is defined as
	\begin{equation}\label{eq:kFGR}
	\kFGR =\frac{\omega_0^3 \mu_{ge}^2}{3\pi\epsilon_0c^3\hbar} 
	\end{equation}
	With these  parameters above, $\kFGR =1.6\times 10^{-3}$.
	To characterize the separation between TLSs, a dimensionless quantity $k_0R = \frac{\omega_0R}{c}$ is used. Finally, we will choose the intermolecular separation to be $k_0R = 0.4$ (by default), corresponding to the dipole-dipole interaction $v_{dd} = \frac{\mu_{ge}^2}{4\pi\epsilon_{0}R^3} = 1.2\times 10^{-2}$. 
	
	Since we operate in vacuum with no dielectric, we can calculate the time-dependent E-field by the dyadic Green's function technique\cite{Novotny2006} instead of numerically solving Eq. \eqref{eq:Maxwell} in a three-dimensional grid\cite{Taflove2005}; see Appendix \ref{Appendix: GreensFunction} for details. We numerically solve the reduced equation of motion for the molecular subsystem by a Runge-Kutta fourth-order propagator\cite{Butcher2008} with the time step $\Delta t = 0.01$.

	\section{Results}\label{sec:result}
	
	After introducing the model Hamiltonian and relevant dynamical methods, we will now perform simulations to mimic two different phenomena: 
	(i) resonance energy transfer (RET) with no external EM field, and (ii) driven dynamics under an external driving cw field.
	 For each case, four semiclassical treatments are considered: (i) Hamiltonian \#I FCI; (ii) Hamiltonian  \#I CIS; (iii) Hamiltonian \#II, and (iv)  the hybrid Hamiltonian. To examine the performance of semiclassical approaches, we will compare them against either the time-dependent perturbative QED result\cite{AkbarSalam2010,Salam2018} or  the results of Lehmberg-Agarwal master equation (LAME)\cite{Lehmberg1970,Agarwal1974} --- the standard quantum approach for describing the dynamics of two-level systems (TLSs) in quantum optics; see Appendix \ref{Appendix: LAME} for details. Note that all LAME results presented below are calculated with FCI.

	\subsection{Resonance Energy Transfer (RET)}
    For RET, no external driven field is considered.  The donor (D) is initialized in a superposition state ($c_g\ket{Dg} + c_e\ket{De}$, where $|c_g|, |c_e| > 0$), the acceptor (A) is initialized in the ground state.
    Here, we choose a superposition state for the donor so that we can initialize a time-dependent current density (and therefore EM field) without invoking any external EM fields. It is well known that Ehrenfest dynamics can depend (unphysically) on the initial state for the donor;  for example, if $c_e = 1$, Ehrenfest dynamics do not predict any spontaneous emission and are completely wrong.
    We consider two regimes: short-time dynamics, from which a RET rate ($\kET$)  can be extracted (see Appendix \ref{Appendix:RET_rate} for details), and long-time dynamics, in which  dissipation effects become important.

	\begin{figure}
		\includegraphics[width=1.0\linewidth]{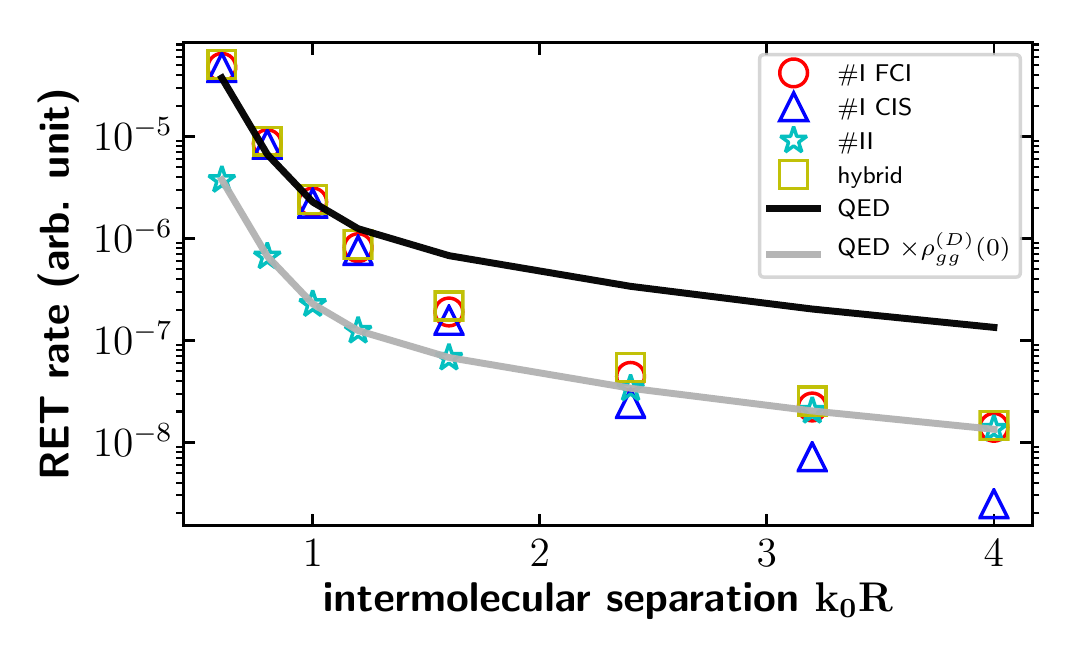}
		\caption{RET rate as a function of intermolecular separation ($k_0R$) according to five approaches: Ehrenfest dynamics with (i) Hamiltonian \#I FCI (red circles), (ii) \#I CIS (blue stars), (iii) \#II (cyan stars), (iv) a hybrid Hamiltonian (yellow squares), and (v) the perturbative QED result (black line). At short range ($k_0R<1$), Hamiltonian \#I and the hybrid Hamiltonian exactly agree with QED due to the use of a quantum dipole-dipole interaction; at long range ($k_0R>1$), no semiclassical approaches can quantitatively predict the QED result because all methods ignore vacuum fluctuations, and the correct physical mechanism is akin to spontaneous emission from one TLS followed by absorption by the other TLS. Note that the RET rate predicted by Hamiltonian \#II is exactly the QED rate times the initial ground state population of the donor [$\rho_{gg}^{(D)}(0)$, the grey line]; see Appendix \ref{Appendix:RET_rate} for an analytic proof. The donor is initialized to $\sqrt{\frac{1}{10}}\ket{Dg} + \sqrt{\frac{9}{10}}\ket{De}$ and the acceptor starts off in the ground state; all other parameters are the same as Ref. \cite{Li2018Tradeoff}.}
		\label{fig:RET_rate}
	\end{figure}
	
	\paragraph{RET rate} Fig. \ref{fig:RET_rate} plots the RET rate as a function of intermolecular separation ($k_0R$, where $k_0 \equiv \frac{\omega_0}{c}$).
	Here, the perturbative QED calculation (black line) suggests that the RET rate obeys two mechanisms in different separation limits:  at short range ($k_0R \ll 1$), the RET rate scales as $\frac{1}{R^6}$ due to  dipole-dipole interactions, known as F\"orster resonance energy transfer (FRET)\cite{Forster1948}; at long range ($k_0R \gg 1$), the RET rate scales as $\frac{1}{R^2}$ because the transverse E-field dominates energy transfer\cite{Andrews1989}. 
	In general, all semiclassical approaches qualitatively predict these scalings but not quantitatively. For example,
	at short range, Hamiltonian \#I [FCI (red circles) and CIS (blue triangles)] and the hybrid Hamiltonian (yellow squares) quantitatively agree with QED while Hamiltonian \#II (cyan stars) predicts only a fraction of the true RET rate [proportional to the ground state population of the donor $\rho_{gg}^{(D)}(0)$; see Appendix \ref{Appendix:RET_rate} for an analytic proof]. 
	 
	 At long range, not surprisingly, because  all semiclassical approaches use a classical E-field and  ignore vacuum fluctuations, none of the methods can predict the RET rate correctly (when $|c_g| \ll 1$). After all, in this limit, the correct physical mechanism is akin to spontaneous emission from one TLS followed by absorption by the other TLS.	Interestingly, in this limit, Hamiltonian \#I CIS predicts an RET rate with a larger error than Hamiltonian  \#I FCI; the underlying reason for this deterioration of accuracy is not obvious because, according to QED,  excluding the doubly excited state should not alter the RET rate if  the double is not populated initially (as is true for RET). 
	 

	\begin{figure*}
		\includegraphics[width=0.8\linewidth]{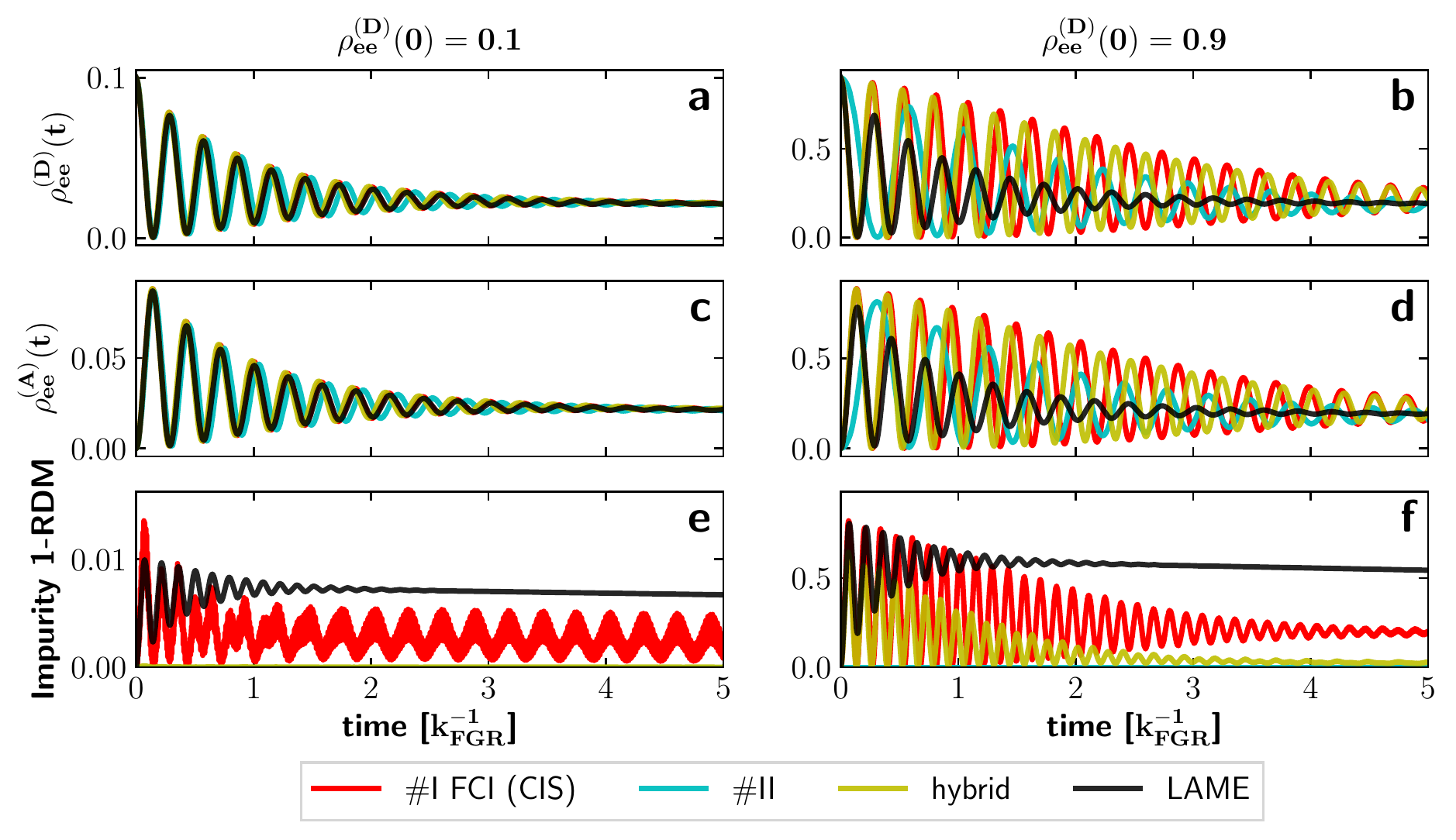}
		\caption{Long-time RET population dynamics as a function of time when $k_0R = 0.4$. (Left) excited state population for the (a) donor and (c) acceptor, and (e) impurity of one-electron reduced density matrix (1-RDM) when $\rho_{ee}^{(D)}(0) = 0.1$; (right) the same dynamics when $\rho_{ee}^{(D)}(0) = 0.9$. Several approaches are compared: Ehrenfest dynamics with (i) Hamiltonian \#I FCI or CIS (these dynamics are identical here and represented by only one single solid red line), (ii)  \#II (solid cyan), (iii) a hybrid Hamiltonian (solid yellow), and (iv) the Lehmberg-Agarwal master equation (LAME,  solid black). Note that all semiclassical approaches agree with LAME when the donor is  weakly excited  initially (see left panel) but predict less dissipation when the donor is strongly excited initially (see right panel).
		All parameters are set to the default values in Sec. \ref{sec:model}.}
		\label{fig:RET_dynamics}
	\end{figure*}

	\begin{figure}
	\includegraphics[width=1.0\linewidth]{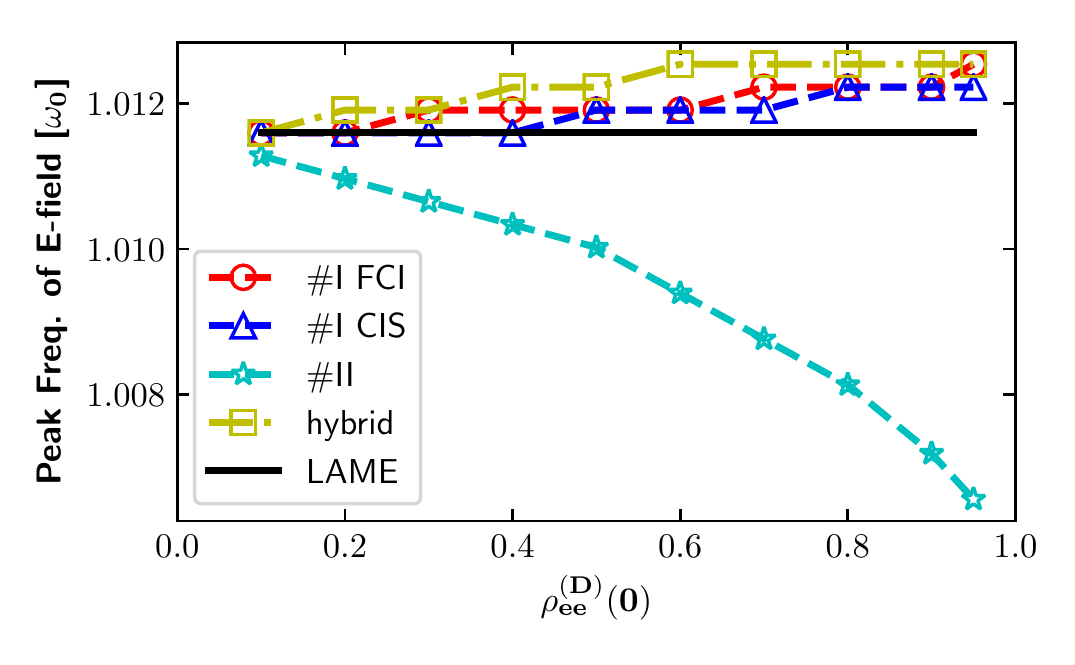}
	\caption{Peak frequency of the scattered E-field as a function of the initial excited state population for the donor [$\rho_{ee}^{(D)}(0)$].
	 Hamiltonian \#II disagrees with  LAME  when $\rho_{ee}^{(D)}(0)$ increases, while the other semiclassical approaches agree with LAME relatively well. A Fourier transform of the scattered E-field is performed when $0 < t < \kFGR^{-1}$, and we choose the frequency with the largest Fourier amplitude. All other parameters are the same as in Fig. \ref{fig:RET_dynamics}.}
	\label{fig:frequency_shift}
	\end{figure}

	\paragraph{Long-time RET dynamics} Fig. \ref{fig:RET_dynamics} plots (from top to bottom) the long-time RET population dynamics for the donor and acceptor, as well as the impurity of the one-electron reduced density matrix (1-RDM)  when the TLSs are close ($k_0R = 0.4$). 
	Here, the impurity of the 1-RDM is a measure to characterize how much the electronic states of different molecules are mixed. For example, when Hamiltonian \#II is used, because the total wavefunction for a pair of TLSs can always be separated as a product of the wavefunctions for each TLS  (which is certainly not true if other approaches are used), the impurity of 1-RDM is always zero (provided it starts at zero). Formally, the impurity of 1-RDM is calculated by $\tr{D} - \tr{D^2}$, where the matrix elements of the 1-RDM ($D$) are defined to be:
	\begin{equation}
	\label{eq:correlation}
	D_{\mu i, \nu j}  = \avg{\Psi_N\Big|\hat{a}^{\dagger}_{i\mu} \hat{a}_{j\nu}\Big|\Psi_N}
	\end{equation} 
	Here, $\left\{ \mu, \nu\right\} = \{1, 2, \cdots, N\}$, $\{i, j\} = \{e, g\}$,  $\hat{a}^{\dagger}_{\mu i}$ and  $\hat{a}_{\mu i}$ are the creation and annihilation operators for state $\ket{\mu i}$. 
	
	When the donor is weakly excited initially [$\rho_{ee}^{(D)}(0)=0.1$; left panel], all semiclassical approaches predict the same population dynamics (Figs. \ref{fig:RET_dynamics}a,c) as the Lehmberg-Agarwal master equation (LAME, black line). 
	These predictions agree with the consensus that a mean-field approximation should be valid when the donor is weakly excited, i.e., in the perturbative regime, where a classical E-field is good enough.
	When the donor is strongly excited [$\rho_{ee}^{(D)}(0)=0.9$;  right panel],  the semiclassical approaches can still predict some key features in population dynamics like oscillations (due to the dipole-dipole interaction), the dissipation, and the long-time slow decay of the dark state, but the dissipation rate is underestimated compared to LAME.
	In general, due to a lack of  quantum dipole-dipole interactions, Hamiltonian \#II (cyan solid) predicts slightly less accurate oscillation periods than other semiclassical approaches.
	More interestingly, for the impurity of the matter 1-RDM (Figs. \ref{fig:RET_dynamics}e,f),  we find that the more one properly accounts for quantum dipole-dipole interactions, the larger is the impurity of the matter subsystem as predicted by semiclassical dynamics  [i.e., as far as the  impurity of the matter subsystem, LAME $>$ Hamiltonian \#I FCI(CIS) $>$ the hybrid Hamiltonian $>$ Hamiltonian \#II $ = 0$]. In Fig. \ref{fig:RET_dynamics}f,  LAME predicts an impurity around $\frac{1}{2}$ at the long times, which can be understood as follows:  for a pair of TLSs in vacuum, if the donor is fully excited, 
	the final state for the TLSs plus the photonic field should be
	  $\frac{1}{\sqrt{2}}\ket{gg;1} + \frac{1}{\sqrt{2}}\ket{d;0}$, where $\ket{gg;1}$ denotes the TLSs in the ground state plus an emitted photon and $\ket{b;0}$ denotes the TLSs in the bright state associated with no photon; thus, the corresponding reduced density matrix for the electronic degrees of freedom is  $\sigma = \begin{psmallmatrix}\frac{1}{2} & 0\\ 0 & \frac{1}{2}\end{psmallmatrix} $, so that the impurity is $\frac{1}{2}$. By contrast, the fact that Ehrenfest is too pure (with an impurity much smaller than LAME) is a statement that additional decoherence is needed.

	Apart from the RET dynamics of the two-level molecules, it is also worthwhile to study the frequency of the scattered E-field.
	 Fig. \ref{fig:frequency_shift} plots the frequency of the scattered E-field as a function of $\rho_{ee}^{(D)}(0)$ for RET dynamics.
	 As predicted by LAME, the frequency of the E-field should not depend on $\rho_{ee}^{(D)}(0)$. However, we find  that Hamiltonian \#II disagrees with LAME  and shows a slightly nonphysical behavior  when $\rho_{ee}^{(D)}(0)$ gradually increases; by contrast,  all other semiclassical approaches agree with LAME relatively well. 
	
	From the above RET results, we gather that Hamiltonian \#II is slightly less accurate  than the other semiclassical approaches especially when the donor becomes more than  weakly excited, in which case one should  include  quantum dipole-dipole interactions.



	\subsection{Collectively Driven Dynamics}
	Now, let us move to the case of collectively driven dynamics for a pair of TLSs prepared initially in ground state. The incident cw field takes the following form: $\vEin(\vR, t) = E_0\sin(\omega_0 t - k_0x) \ve_z$. To characterize the strength of the cw field, the Rabi frequency ($\Omega \equiv \mu_{ge}E_0$) is a good indicator: $\Omega < \kFGR$ ($\Omega > \kFGR$) represents a weak (strong) driving field.
	In general, for closely aggregated TLSs ($k_0 R \ll 1$), because the spontaneous emission rate is strongly modified by intermolecular interactions ($v_{dd}$) instead of the vacuum value in Eq. \eqref{eq:kFGR}, one would expect that semiclassical approaches should be valid as long as the Rabi frequency is much smaller than the dipole-dipole coupling ($\Omega \ll v_{dd}$). 
	 With this in mind, we check the results of driven dynamics as below.

		\begin{figure}
		\includegraphics[width=1.0\linewidth]{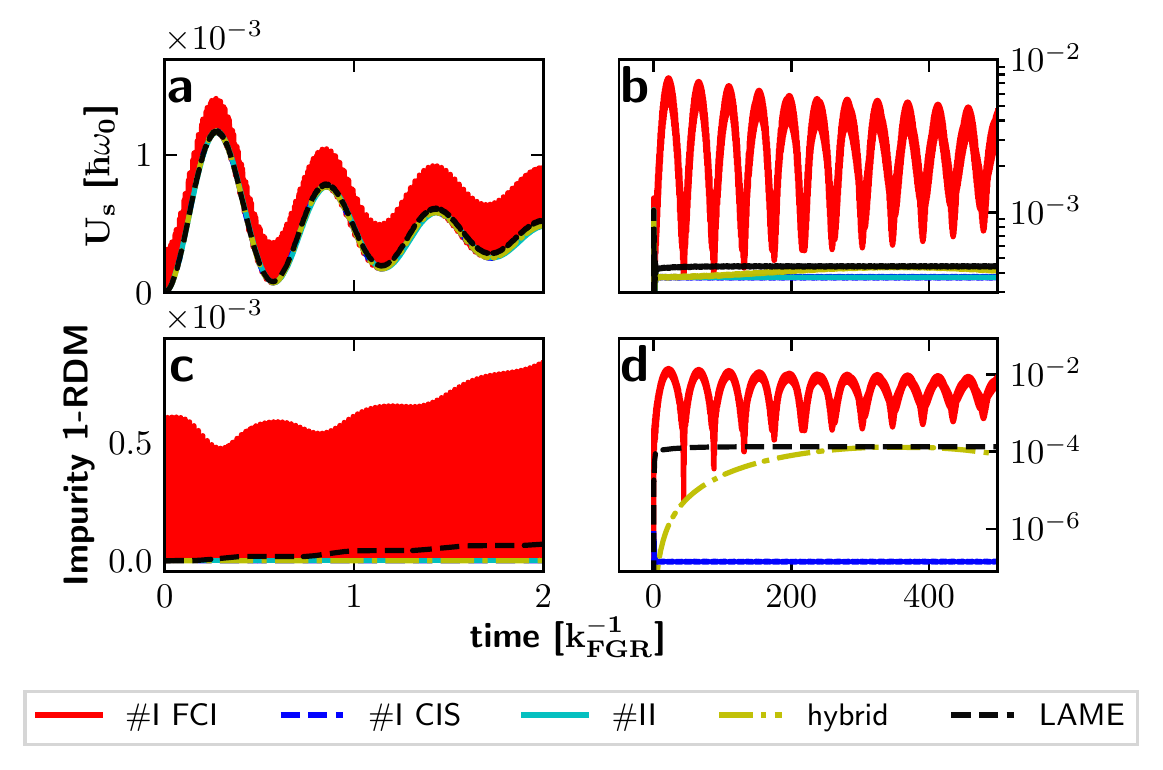}
		\caption{Electronic energy (upper) and impurity of the matter 1-RDM (bottom) for a pair of TLSs as a function of time driven by a weak cw field ($\Omega \equiv \mu_{ge}E_0 = 0.3\kFGR$): (left) the early dynamics ($t < 2\kFGR^{-1}$); (right) the steady-state dynamics ($t \sim 300\kFGR^{-1}$; logarithmic scale for y-axis).  Note that all approaches predict similar dynamics for electronic energy expect that in steady state, Hamiltonian \#I FCI predicts an unphysically large electronic energy; see Fig. b. All parameters are set as the default values in Sec. \ref{sec:model}.}
		\label{fig:driven_weak}
	\end{figure}

	\paragraph{Weakly driven dynamics} Fig. \ref{fig:driven_weak} plots the electronic energy (Fig. \ref{fig:driven_weak}a-b)  and the impurity of 1-RDM  (Fig. \ref{fig:driven_weak}c-d)  for  a pair of TLSs driven by a weak cw field ($\Omega = 0.3\kFGR$)
	at both short times (left panel) and long times (right panel). Here, 
	the electronic energy of the molecular subsystem ($\Us$) is defined as
	\begin{equation}
	\label{eq:Us}
	\Us = \sum_{n=1}^{N}\tr{\hrho(t) \hH_s^{(n)}}
	\end{equation}
	As explained above, we expect that all approaches (Hamiltonian \#I FCI, \#I CIS, \#II, the hybrid Hamiltonian, and LAME) should predict the same dynamics for electronic energy. The surprising finding, however, is that after very long times ($t > 200\kFGR^{-1}$), Hamiltonian \#I FCI (red solid) predicts an unphysically large electronic energy compared to other approaches; see Fig. \ref{fig:driven_weak}b. This unphysical behavior  indicates (ironically) that a full accounting for quantum electron-electron correlations can actually be problematic even in the weak coupling limit. The reason for this anomaly will be addressed in Sec. \ref{sec:discussion}. For the impurity of the 1-RDM, as shown in Fig. \ref{fig:driven_weak}d, while Hamiltonian \#I FCI overestimates the impurity as compared with LAME, the hybrid Hamiltonian predicts similar steady-state impurity as LAME, and other semiclassical approaches predict much less impurity than LAME (note that here Hamiltonian \#II still always predicts zero impurity).

	\begin{figure}
		\includegraphics[width=1.0\linewidth]{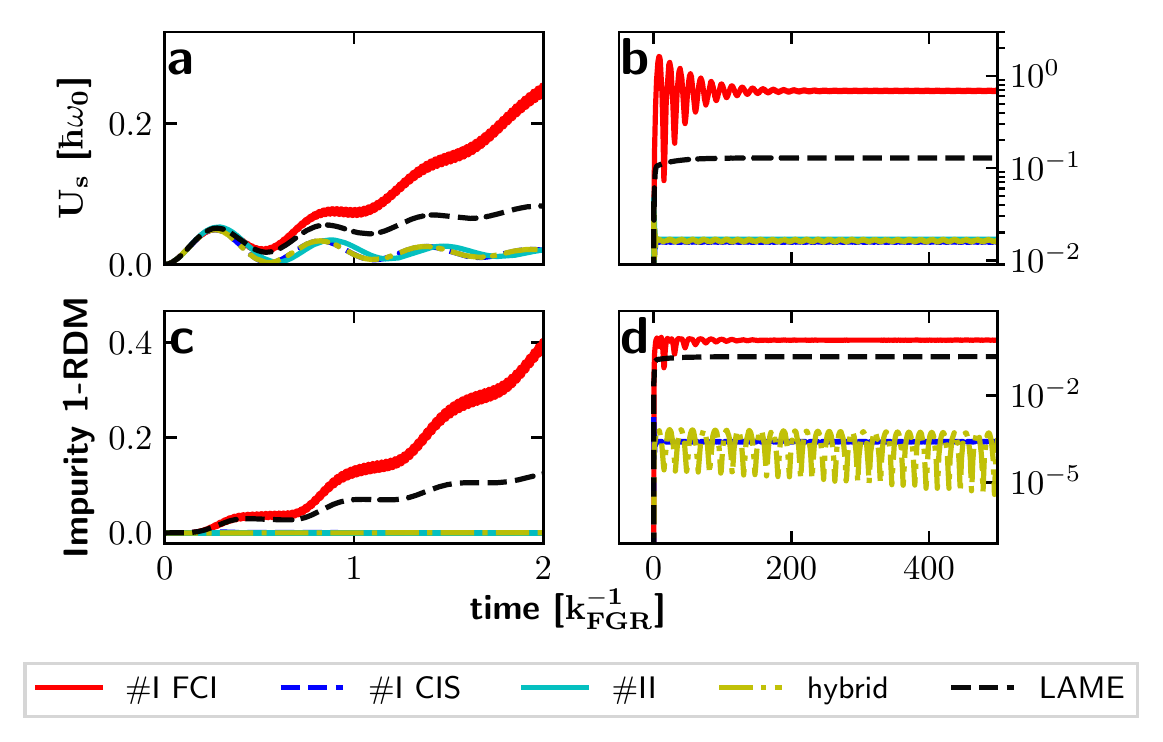}
		\caption{ The same plot as Fig. \ref{fig:driven_weak} but with a strong cw wave ($\Omega = 2.0\kFGR$).
			Note that because the contribution of the double is significant under a strong driving field, LAME predicts more electronic energy $\Us$) than does Hamiltonian  \#I CIS, for which the  double is truncated. Perhaps surprisingly, Hamiltonians \#II and the hybrid Hamiltonian predict similar behavior of electronic energy as Hamiltonian \#I CIS (even though the former has the capacity to describe the double). As in Fig. \ref{fig:driven_weak}, Hamiltonian \#I FCI still greatly overestimates the electronic energy and impurity.}
		\label{fig:driven_strong}
	\end{figure}

	\paragraph{Strongly driven dynamics}
	Fig. \ref{fig:driven_strong} plots the dynamics of the electronic energy and the impurity of 1-RDM when the cw field becomes stronger ($\Omega =  2.0\kFGR < v_{dd}$). In this limit, because the contribution of the double is not negligible, as is shown in Fig. \ref{fig:driven_strong}b(d), LAME predicts a much higher steady-state electronic energy  (and impurity) than does Hamiltonian \#I CIS, for which the double is truncated.
	Just as in Fig. \ref{fig:driven_weak}, by including the double, Hamiltonian \#I FCI  overestimates the electronic energy significantly compared with LAME, reinforcing the notion that fully accounting for electron-electron correlation can be problematic (in both the  weak and strong field limits).
	As far as the impurity of the matter 1-RDM (Fig. \ref{fig:driven_strong}c,d.), the behaviors of the different semiclassical approaches are similar to what was found in the case of electronic energy, except for the fact that Hamiltonian \#II always predicts zero impurity.
	Apparently, electronic FCI coupled to  a classical EM field can predict nonphysical features, which conflicts with our intuition that including more electron-electron correlations should give better results. 
	
	Overall, for a reasonably strong field, no semiclassical approach can predict the steady-state electronic energy or the impurity of the matter 1-RDM correctly, which would naively conflict with the general consensus that semiclassical electrodynamics should be valid as long as the Rabi frequency ($\Omega$) is much smaller than the strength of the dipole-dipole coupling ($v_{dd}$). The validity of semiclassical electrodynamics is obviously complicated, and must depend on which Hamiltonian one uses. With this in mind, let us now digest the results above and consider why FCI behaves so poorly in Figs. \ref{fig:driven_weak} and \ref{fig:driven_strong}.
	
	\section{Discussion}
	\label{sec:discussion}
	
	From the results above in Figs. \ref{fig:RET_rate}-\ref{fig:driven_strong}, our general conclusion is that no semiclassical method is perfect, but Hamiltonian \#I CIS and the hybrid Hamiltonian seem to perform optimally and they are reasonably computationally efficient. Hamiltonian \#II performs slightly worse (failing for RET and the impurity of 1-RDM). The most stunning conclusion is the drastic failure of Hamiltonian \#I FCI under driven dynamics.
		
\begin{figure}
		\includegraphics[width=1.0\linewidth]{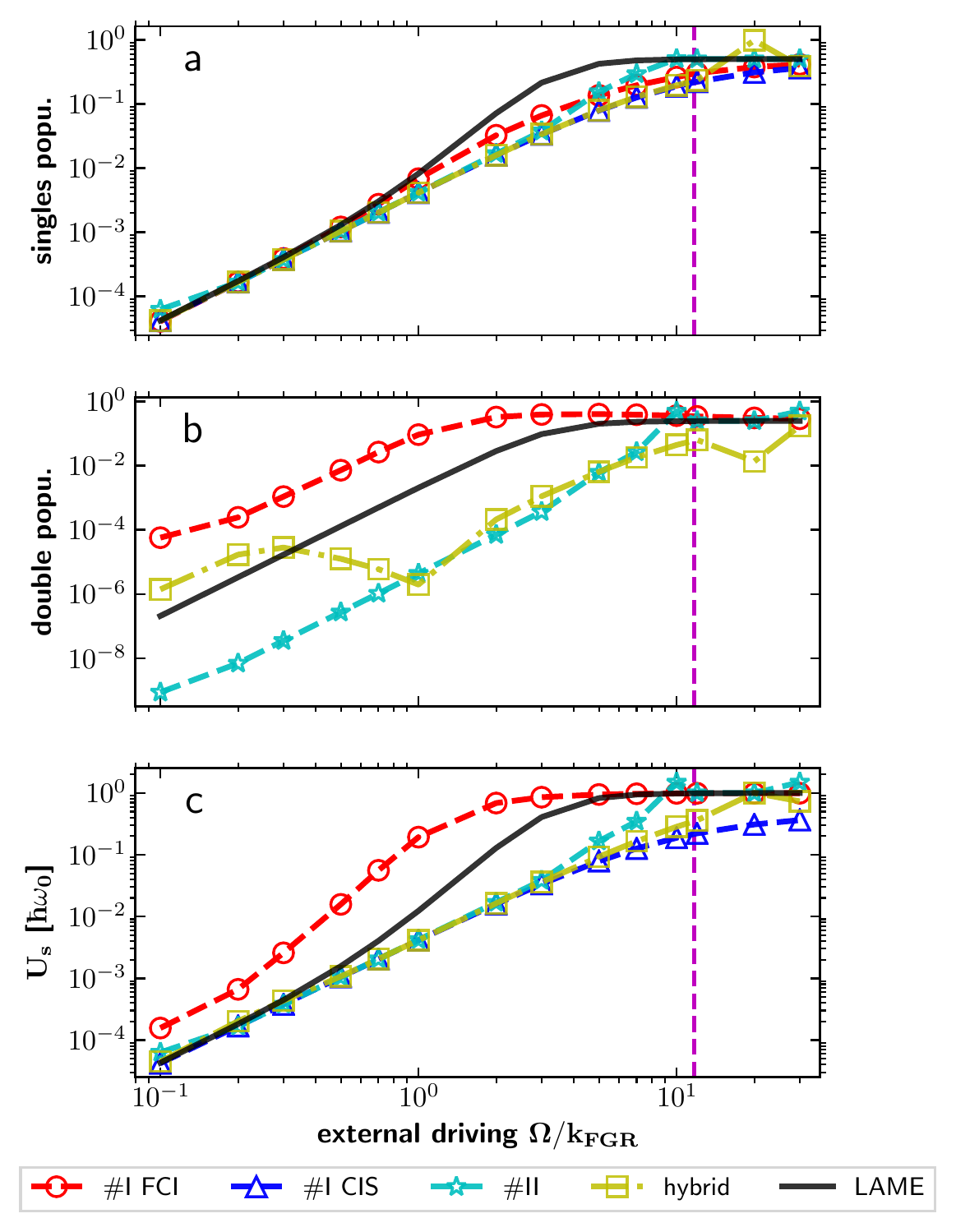}
		\caption{ Plots of the steady-state (a) population of singles, (b) population of the double, and (c) electronic energy as a function of the external driving strength ($\Omega / \kFGR$) on a logarithmic scale. The  vertical magenta line denotes $\Omega = v_{dd}$. Fig. \ref{fig:ratios}a shows that when $\Omega \ll v_{dd}$, all semiclassical approaches predict similar values for the single populations as LAME does. Fig. \ref{fig:ratios}b shows that \#I FCI (\#II) overestimates (underestimates) the population of the double greatly even when $\Omega \ll v_{dd}$. }
		\label{fig:ratios}
	\end{figure}

	To better understand the failure of FCI in driven dynamics, consider the steady-state data in Fig. \ref{fig:ratios}a. When the Rabi frequency ($\Omega$, $x$-axis) is much smaller than the dipole-dipole coupling ($v_{dd}$, the vertical magenta line), all semiclassical approaches predict similar steady-state population for the singles ($y$-axis) as compared to LAME. 
	However, when we investigate  the  population of the double excitation (see Fig. \ref{fig:ratios}b), conventional semiclassical approaches fail even when $\Omega \ll v_{dd}$. 
	On the one hand, Hamiltonian \#II always greatly underestimates the population of the double. If we restrict ourselves to the weak coupling limit ($\Omega \ll \kFGR$), such an underestimation is not very problematic because the population is so small as to have minimal effect on any physical observables.
	On the other hand, Hamiltonian \#I FCI always  overestimates the population for the double, leading to a nonphysically large electronic energy; see Fig. \ref{fig:ratios}c. One may therefore hypothesize that the inclusion of the doubly excited state represents an important but risky proposal for semiclassical electrodynamics; overestimation of the double population is strongly correlated to the overestimation of the total electronic energy.
	Interestingly, the hybrid Hamiltonian does interpolate between Hamiltonian \#I and \#II, but there is minimal gain in accuracy when $\Omega \gg \kFGR$.

	We can now answer the question above: why does FCI fail and predict an exorbitant accumulation of energy for the TLSs under a driving force? The root of this problem is the classical EM field.
	Note that, for a single TLS, due to the use of a classical EM field, Ehrenfest dynamics predicts a decay rate proportional to the ground state population\cite{Crisp1969,Milonni1976,Li2018Spontaneous}:
	\begin{equation}
		\label{eq:Eh_rate}
		\kEh = \rho_{gg}\kFGR
	\end{equation}
	For a pair of closely aggregated TLSs ($k_0R \ll 1$, as considered in this manuscript), if one neglects the effect of the dark state and focuses on a  three-level system with ground state $\ket{0}$, bright state $\ket{b}$, and doubly excited state $\ket{2}$,  the allowed optical transitions are $\ket{0} \leftrightarrow \ket{b}$ and $\ket{b} \leftrightarrow \ket{2}$ [and the Ehrenfest decay rates between these optical transitions also obey Eq. \eqref{eq:Eh_rate}]. 
	 For driven dynamics, with system initially in state $\ket{0}$, the quantum dipole-dipole interaction $\hV_{\text{Coul}}^{(nl)}$ directly couples state $\ket{0}$ and state $\ket{2}$. Now, suppose we apply Hamiltonian \#I with FCI.  On the one hand, with driven dynamics,  $\hV_{\text{Coul}}^{(nl)}$ leads to an increase of the population for state $\ket{2}$; on the other hand, because initially $\rho_{bb}(0) = 0$, according to Eq. \eqref{eq:Eh_rate}, the decay rate from $\ket{2}$ to $\ket{b}$ is greatly suppressed. As a result, state $\ket{2}$ will continuously accumulate the population, leading to an unphysically large electronic energy even in the weak coupling limit. 
	In short, the exaggerated electronic energy predicted by Hamiltonian \#I FCI (see Figs. \ref{fig:driven_weak}-\ref{fig:ratios}) appears to come directly from the mismatch of the quantum electron-electron correlations and the classical EM field. Interestingly, this mismatch also causes the violation of long-range causality\cite{Li2018Tradeoff}.
	The above discussion should be very general, valid for a pair of TLSs or in the case of many molecules: for driven systems, the population dynamics for higher excitations (beyond singles) cannot be correctly described by Hamiltonian \#I FCI even when the driving field is very weak. 
	

	\section{Conclusion}\label{sec:conclusion}
	To conclude, in this manuscript, we have applied different semiclassical approaches to a minimalistic many-site model for light-matter interactions --- a pair of identical TLSs. We find: 
	(i) For the impurity of the 1-RDM, generally no semiclassical approach agrees with LAME very well;
	(ii) For RET dynamics, Hamiltonian \#II is not an optimal candidate due to a lack of  quantum dipole-dipole couplings;
	(iii) For collectively driven dynamics, all semiclassical approaches in Table \ref{table:Hamiltonians} can correctly describe the population of single states when the Rabi frequency is much smaller than the dipole-dipole coupling ($\Omega \ll v_{dd}$);
	(iv) For collectively driven dynamics, even when $\Omega \ll v_{dd}$,
	Hamiltonian \#I FCI always predicts a nonphysically large double population  (and thus an incorrect electronic energy) due to a mismatch between quantum electron-electron correlations and a classical E-field;
	(v) A hybrid Hamiltonian can eliminate the reported anomaly for \#I FCI in the weak field as well as outperform Hamiltonian \#II  with regard to RET. Nevertheless, the accuracy of the hybrid Hamiltonian is still far from quantitative.

	For the moment, when using semiclassical electrodynamics to describe light-matter interactions, our recommendation is to use Hamiltonian \#I CIS or the hybrid Hamiltonian as a trade-off between accuracy and computational cost.
	We must emphasize that (i) our present benchmark work was restricted to only a pair of TLSs, and (ii) no semiclassical algorithm performs quantitatively at all.
	In the future, these limitations must be addressed.
	On the one hand, for a large collection of molecules, more exciting collective phenomena should emerge and the performances of the different semiclassical approaches must be tested. On the other hand, and even  more importantly, it is also natural to ask whether or not further algorithmic improvements can be made to the semiclassical methods above. For example, can we include some crucial aspects of spontaneous emission that are missed in a mean-field treatment and improve Hamiltonian \#II? Recent experience \cite{Chen2018Spontaneous} suggests such improvements are possible and this work in ongoing.

	\section{Acknowledgement}\label{sec:acknowledgement}
	This material is based upon work supported by the U.S. Department of Energy, Office of Science, Office of Basic Energy Sciences under Award Number DE-SC0019397.
	The research of A.N. is supported by the Israel-U.S. Binational
	Science Foundation.
	This research also used resources of the National Energy Research Scientific Computing Center (NERSC), a U.S. Department of Energy Office of Science User Facility operated under Contract No. DE-AC02-05CH11231.

	\begin{appendices}

		\section{Analytical and EM-Free Form of Semiclassical Hamiltonians}
		\label{Appendix: GreensFunction}
	
		\setcounter{equation}{0}
		\setcounter{figure}{0}
		\setcounter{table}{0}
		\renewcommand{\theequation}{A\arabic{equation}}
		\renewcommand{\thefigure}{A\arabic{figure}}
		
		\subsection{Longitudinal and Transverse Components}
		For a vector function $\mathbf{f}(\vR) = f_x(\vR)\ve_x + f_y(\vR)\ve_y + f_z(\vR)\ve_z$, the longitudinal component is defined by
		\begin{equation}
			\label{eq:longitutional-transverse-compact}
			\begin{aligned}
				\mathbf{f}_{\parallel}(\vR) = \int d\vR' \overleftrightarrow{\vdelta}_{\parallel}(\vR - \vR') \mathbf{f}(\vR')
			\end{aligned}
		\end{equation}
		where the dyadic longitudinal $\delta$-function $\overleftrightarrow{\vdelta}_{\parallel}(\vR)$ is 
		\begin{equation}
			\label{eq:dyadic-longitutional-delta-function}
			\overleftrightarrow{\vdelta}_{\parallel}(\vR) = \sum\limits_{i,j=x,y,z} \delta_{\parallel ij}(\vR) \he_i\he_j
		\end{equation}
		Here, $\he_i$ denotes a unit vector along direction $i = x, y, z$, and 
		\begin{subequations}
			\label{eq:longitutional-delta-function}
			\begin{align}
				\label{eq:longitutional-1}
			\delta_{\parallel i j}(\mathbf{r}) &= - \nabla_i\nabla_j \frac{1}{4\pi |\vR|}\\
				\label{eq:longitutional-2}
			&=\frac{1}{3} \delta_{i j} \delta(\mathbf{r}) - \frac{\eta(\mathbf{r})}{4 \pi |\vR|^{3}}
							\left(\frac{3 r_{i} r_{j}}{|\vR|^{2}}-\delta_{i j}\right) 		
			\end{align}
		\end{subequations}
		While the first definition, Eq. \eqref{eq:longitutional-1}, is a natural definition of the longitudinal $\delta$-function, this expansion diverges at $|\vR| = 0$. To avoid such divergence, regularization is introduced, leading to the second definition, Eq. \eqref{eq:longitutional-2}, in which \(\eta(\mathbf{r})\equiv0\) at \(\mathbf{r}=0\) to suppress the divergence and \(\eta(\mathbf{r})\equiv1\) elsewhere\cite{Cohen-Tannoudji1997}. 

		Similar to Eq. \eqref{eq:dyadic-longitutional-delta-function}, the dyadic transverse $\delta$-function  reads
		\begin{equation}
			\label{eq:transverse-delta-function}
			\overleftrightarrow{\vdelta}_{\perp}(\vR) = \sum\limits_{i,j=x,y,z} \delta_{\perp ij}(\vR) \he_i\he_j
		\end{equation}
		Note that $\delta_{\perp i j}(\mathbf{r}) \equiv \delta_{ i j}(\mathbf{r}) - \delta_{\parallel i j}(\mathbf{r})$, so that the transverse component $\mathbf{f}_{\perp}(\vR)$ can be calculated by 
		\begin{equation}
		\label{eq:longitutional-transverse-compact-2}
		\mathbf{f}_{\perp}(\vR) = \int d\vR' \overleftrightarrow{\vdelta}_{\perp}(\vR - \vR') \mathbf{f}(\vR')
		\end{equation}
		According to the definitions of the longitudinal and transverse $\delta$-functions, it is easy to show that $\int d\vR \ \mathbf{f}_{\perp}(\vR) \cdot \mathbf{f}_{\parallel}(\vR) = 0$ for all vector fields $\mathbf{f}(\vR)$.

		\subsection{Time-Dependent Dyadic Green's Functions}
			
		If we assume that the  electronic subsystem  couples only to the E-field (as is true in this manuscript), it is more convenient to rewrite Maxwell's equations [Eq. \eqref{eq:Maxwell}] as
		\begin{equation}
			\label{eq:Maxwell-Eform}
			\vnabla\times\vnabla\times \vE(\vR, t) + \frac{1}{c^2}\frac{\partial^2 \vE(\vR, t) }{\partial t^2} = - \mu_0 \sum_{n}
			\frac{\partial^2 \vP^{(n)}(\vR, t)}{\partial t^2}
		\end{equation} 
		A formal solution of the E-field reads
		\begin{equation}
			\label{eq:E-field-formal}
			\vE(\vR, t) = \vEin(\vR, t) + \sum_{n}\vE^{(n)}(\vR, t)
		\end{equation}
		where $\vEin(\vR, t)$ denotes the incoming field, and $\vE^{(n)}(\vR, t)$ denotes the E-field that is emitted by molecule $n$, which 
		can be further evaluated through the time-dependent dyadic Green's function technique\cite{Novotny2006}, i.e.,
		\begin{equation}
			\label{eq:E-field-solution-Greens}
			\vE^{(n)}(\vR, t) = \mu_0 \omega^2 \int_V dV'\int  dt'\vGreen(\vR, \vR'; t, t')\vP^{(n)}(\vR', t') 
		\end{equation}
		where $V$ denotes the integral volume that includes $\vP^{(n)}$. The time-dependent dyadic Green's function $\vGreen(\vR, \vR', t, t')$ is defined as
		\begin{equation}
			\label{eq:Green-function}
			\vGreen(\vR, \vR'; t, t') = \left[ \vI + \frac{1}{k^2}\vnabla\vnabla \right]G_0(\vR, \vR'; t, t')
		\end{equation}
		where $k = \frac{\omega}{c}$.
		For a point source in a homogeneous environment, the time-dependent scalar Green's function $G_0$ reads
		\begin{equation}
			\label{eq:scalar-Green-function}
			G_0(\vR, \vR'; t, t') = \frac{\delta\left( t' - \left[ t - \frac{n}{c}|\vR - \vR'| \right] \right)}{4\pi|\vR - \vR'|}
		\end{equation}
		where $n = 1$ in vacuum. By substituting Eqs. \eqref{eq:Green-function} and \eqref{eq:scalar-Green-function} into Eq. \eqref{eq:E-field-solution-Greens}, we arrive at a retarded expression of $\vE^{(n)}$:
		\begin{widetext}
		\begin{equation}
			\label{eq:E-field-Greens-retarded}
			\begin{aligned}
						\vE^{(n)}(\vR, t)=  \mu_0 \omega^2  \int_V dV' \left[ \vI + \frac{1}{k^2}\vnabla\vnabla
			\right]  \frac{\vP^{(n)}\left ( \vR', t - \frac{n}{c}|\vR - \vR'| \right ) }{4\pi|\vR - \vR'|}
			\end{aligned}
		\end{equation}
		\end{widetext}
	
		Now, very often, within the content of electrodynamics with retardation, it is helpful to work with the time-independent dyadic Green's function $\vGreen(\vR, \vR')$: 
		\begin{equation}
			\label{eq:greens-functions-short-hand}
			\begin{aligned}
				\vGreen(\vR, \vR') &= \left[ \vI + \frac{1}{k^2}\vnabla\vnabla \right]G^{-}_0(\vR, \vR') \\
				G^{-}_0(\vR, \vR')  &= \frac{e^{-ik |\vR - \vR'|}}{4\pi |\vR - \vR'|} = \frac{e^{-ik R}}{4\pi R}
			\end{aligned}
		\end{equation}
		where $R \equiv |\vR - \vR'|$. Eq. \eqref{eq:greens-functions-short-hand} can be rewritten as
		\begin{equation}
			\label{eq:greens-functions-expression}
			\vGreen(\vR, \vR') = \frac{e^{-ikR}}{4\pi R} \left[ 
			\veta_1  - \frac{i}{kR} \veta_3 -\frac{1}{k^2 R^2}\veta_3 \right]
		\end{equation}
		where $\veta_1$ and $\veta_3$ are defined as
		\begin{subequations}
			\label{eq:eta}
			\begin{align}
				\label{eq:eta1}
				\veta_1 &= \vI - \hat{\vRR}_i\hat{\vRR}_j \\
				\label{eq:eta2}
				\veta_3 &= \vI - 3\hat{\vRR}_i\hat{\vRR}_j
			\end{align}
		\end{subequations}
		and $\hat{\vRR}_i$ denotes the unit vector along the direction of $\vRR_i = \vR_i - \vR_i'$.
		
		Because we will have different molecules at different sites, let us also introduce the following short-hand writing
		\begin{equation}
		\label{eq:Greens-short-hand}
		\begin{aligned}
		G_{nl} & \equiv \ve_d^{(n)}\cdot\vGreen(\vR_n, \vR_l)\ve_d^{(l)} \\
		& = \frac{e^{-ikR_{nl}}}{4\pi R_{nl}} \left[ 
		\eta_1^{(nl)}  - \frac{i}{kR_{nl}} \eta_3^{(nl)} -\frac{1}{k^2 R_{nl}^2}\eta_3^{(nl)} \right]
		\end{aligned}
		\end{equation}
		where $R_{nl} \equiv |\vR_n - \vR_l|$, $ \ve_d^{(n)}$ denotes the unit vector along the orientation of dipole $n$,  $\eta_1^{(nl)} = \ve_d^{(n)}\cdot \veta_1\ve_d^{(l)}$ and $\eta_3^{(nl)} = \ve_d^{(n)}\cdot \veta_3\ve_d^{(l)}$.
		$G_{nl}$ in Eq. \eqref{eq:Greens-short-hand} characterizes the magnitude of the light-matter coupling between the two unit dipoles at sites $n$ and $l$. The real and imaginary parts of $G_{nl}$ read
		\begin{widetext}
		\begin{equation}
		\label{eq:greens-functions-expression-2}
		\begin{aligned}
		\real{ G_{nl} } &= \frac{k}{4\pi} \left[  \frac{\cos(kR_{nl})}{kR_{nl}} \eta_1^{(nl)} - \frac{\sin(kR_{nl})}{k^2R_{nl}^2} \eta_3^{(nl)} 
		- \frac{\cos(kR_{nl})}{k^3R_{nl}^3} \eta_3^{(nl)}  \right]\\
		\imag{ G_{nl} } &= \frac{k}{4\pi} \left[  -\frac{\sin(kR_{nl})}{kR_{nl}} \eta_1^{(nl)} - \frac{\cos(kR_{nl})}{k^2R_{nl}^2} \eta_3^{(nl)} 
		+ \frac{\sin(kR_{nl})}{k^3R_{nl}^3} \eta_3^{(nl)} \right]
		\end{aligned}
		\end{equation}
	\end{widetext}
		Interestingly, when two dipoles overlap, i.e., $\ve_d^{(n)} =  \ve_d^{(l)}$, $\eta_1^{(nl)} = \eta_3^{(nl)} = 1$, and $R_{nl} \rightarrow 0$, a Taylor expansion of Eq. \eqref{eq:greens-functions-expression-2} to leading order in $kR$ reduces to
		\begin{equation}
		\label{eq:real-imag-parts}
		\begin{aligned}
		\frac{4\pi}{k} \real{ G_{nn} }\Big|_{R_{nl}\rightarrow 0} &\rightarrow  -  \frac{1}{k^3R^3}\Big|_{R_{nl}\rightarrow 0} \\
		\frac{4\pi}{k}  \imag{ G_{nn} } \Big|_{R_{nl}\rightarrow 0}& \rightarrow  -  \frac{2}{3}
		\end{aligned}
		\end{equation}

		\subsection{Analytical and EM-Free Form of Hamiltonian \#I}
		For Hamiltonian \#I [defined in Eq. \eqref{eq:H-I}], it is unnecessary to evaluate $\vE_{\perp}$ at all times. Instead, for neutral molecules (with no free charge), since the displacement field ($\vD$) is transverse, i.e., $\vD_{\parallel} = \vE_{\parallel} + \frac{1}{\epsilon_0}\vP_{\parallel} = \mathbf{0}$, $\vE_{\perp}$ can be rewritten as
		\begin{equation}
		\label{eq:Eperp-expand}
		\vE_{\perp} = \vE - \vE_{\parallel} = \vE + \frac{1}{\epsilon_0} \vP_{\parallel}
		\end{equation}
		By substituting Eq. \eqref{eq:Eperp-expand} into Eq. \eqref{eq:H-I}, one obtains another form for Hamiltonian \#I:
		\begin{equation}
		\label{eq:H-I-2}
		\begin{aligned}
				\hH_{sc}^{I} = & \sum_{n = 1}^{N}\hH_s^{(n)} -\int d\vR \ \vE(\vR,t)\cdot \hP^{(n)}(\vR)  
		+ \sum_{n < l} \hV_{\text{Coul}}^{(nl)} \\ 
		& - \sum_{nl} \frac{1}{\epsilon_0}\int d\vR \ \vP_{\parallel}^{(l)}(\vR,t)\cdot \hP^{(n)}(\vR)  
		\end{aligned}
		\end{equation}
		At this point, let us evaluate all of the terms in Eq. \eqref{eq:H-I-2}. If we make the long wave approximation, i.e., $\hP^{(n)}(\vR) = \hmu^{(n)} \delta(\vR - \vR_n) \ve_d^{(n)}$ (where $\hmu^{(n)} \equiv \mu_{ge}\ve_d^{(n)}\hsigma_x^{(n)}$ denotes the transition dipole operator for TLS $n$), and apply Eqs. \eqref{eq:longitutional-transverse-compact}-\eqref{eq:longitutional-delta-function}, $\hV_{\text{Coul}}^{(nl)}$ [Eq. \eqref{eq:vdd_QED}] is reduced to the dipole-dipole interaction form:
		\begin{equation}
		\label{eq:Vdd-point-dipoles}
		\begin{aligned}
		\hat{V}_{\mathrm{Coul}}^{(nl)} &=\frac{1}{4 \pi \epsilon_{0}}\left(\frac{\hat{\boldsymbol{\mu}}^{(n)} \cdot \hat{\boldsymbol{\mu}}^{(l)}}{\left|\mathbf{r}\right|^3}-\frac{3\left(\hat{\boldsymbol{\mu}}^{(n)} \cdot \mathbf{r}\right)\left(\hat{\boldsymbol{\mu}}^{(l)} \cdot \mathbf{r}\right)}{|\mathbf{r}|^{5}}\right) \\
		&= 	\frac{ \mu_{ge}^2 \eta_{3}^{(nl)}}{4\pi \epsilon_0 R_{nl}^3}
		\hsigma_x^{(n)}\otimes \hsigma_x^{(l)}
		\end{aligned}
		\end{equation}
		where $\otimes$ denotes the Kronecker tensor product.
		Similarly, for $n\neq l$, the last term in Eq. \eqref{eq:H-I-2} can be simplified as
		\begin{equation}
		\label{eq:intPP-point-dipoles}
		\begin{aligned}
		\hat{v}^{(nl)}_{\text{Coul}}(t) &\equiv \frac{1}{\epsilon_0}\int d\vR \ \vP_{\parallel}^{(l)}(\vR,t)\cdot \hP^{(n)}(\vR)  \\
		&=
		\frac{1}{4 \pi \epsilon_{0}}\left(\frac{\hmu^{(n)} \cdot \vmu^{(l)}(t)}{\left|\mathbf{r}\right|^3}-\frac{3\left(\hmu^{(n)} \cdot \vR\right)\left(\vmu^{(l)}(t) \cdot \vR\right)}{|\mathbf{r}|^{5}}\right) \\
				&=  \frac{2\real{\rho_{ge}^{(l)}} \mu_{ge}^2 \eta_{3}^{(nl)}}{4\pi \epsilon_0 R_{nl}^3}
		\hsigma_x^{(n)}
		\end{aligned} 
		\end{equation}
		where $\vmu^{(l)}(t) \equiv \tr{\hrho(t)\hmu^{(l)}} = 2\real{\rho_{ge}^{(n)}(t)}\mu_{ge}\ve_d^{(n)}$, and  $\rho_{ge}^{(n)}(t)$ denotes the coherence between the ground state and excited state for TLS $n$. For Hamiltonian \#I, 	we can calculate $\rho_{ge}^{(n)}$ by $\rho_{ge}^{(n)}(t) = \tr{\hrho(t)\hsigma_{+}^{(n)}}$.

		At this point, having evaluated all electronic matrix elements in Eq. \eqref{eq:H-I-2}, for  the sake of simplicity and efficiency, we would like to completely reduce Hamiltonian \#I (when possible) into a Hamiltonian operating only on the electronic degrees of freedom, from which the electric and magnetic  fields can be  extrapolated analytically; this is, after all, the framework of the famous optical Bloch equation (OBE).
		To do so, let us evaluate the E-field using a Green's function technique.
		For a TLS under the long wavelength approximation, $\vP^{(n)}(\vR, t)$ reads
		\begin{subequations}
			\label{eq:vPn-TLS-point}
			\begin{align}
			\vP^{(n)}(\vR, t) &= \tr{\hrho(t) \hmu^{(n)}}\delta(\vR - \vR_n)\ve_d^{(n)} \\
			&= 2\mu_{ge}\real{\rho_{ge}^{(n)}(t)}\delta(\vR - \vR_n)\ve_d^{(n)}
			\end{align}
		\end{subequations}
		By substituting Eq. \eqref{eq:vPn-TLS-point} into Eq. \eqref{eq:Green-function}, we arrive at an analytical form for $\vE^{(n)}(\vR, t)$:
		\begin{widetext}
		\begin{subequations}
			\label{eq:E-2TLS}
			\begin{align}
			\label{eq:E-2TLS-1}
			\vE^{(n)}(\vR, t) &= \mu_0\omega^2 \int_V d V' \left[ \vI + \frac{1}{k^2}\vnabla\vnabla \right]
			\frac{2\real{\rho_{ge}^{(n)}\left( t - \frac{|\vR - \vR'|}{c} \right) } }{4\pi |\vR - \vR'|} \\
			\label{eq:E-2TLS-2}
			& = 2\mu_0 \real{\omega^2\rho_{ge}^{(n)}(t)  \vGreen(\vR, \vR_n)\vmu^{(n)} }_{\omega = \omega_{0}} 
			\end{align}
		\end{subequations}
	\end{widetext}
		Between Eq. \eqref{eq:E-2TLS-1} and Eq. \eqref{eq:E-2TLS-2}, we have neglected all retardation and assumed
		\begin{equation}
		\label{eq:rho_ge_retarded}
		\rho_{ge}^{(n)}\left( t - \frac{|\vR - \vR'|}{c} \right)  \approx \rho_{ge}^{(n)}(t)e^{-i \omega_{0}\frac{|\vR - \vR'|}{c}};
		\end{equation}
		 the time-independent Green's function $\vGreen(\vR, \vR_n)$ is defined in Eq. \eqref{eq:greens-functions-short-hand}. Given Eq. \eqref{eq:E-2TLS}, the coupling between molecule $n$ and the E-field generated by molecule $l$ ($n\neq l$) is expressed as 
			\begin{widetext}
		\begin{subequations}
				\label{eq:Omega_nl}
			\begin{align}
			\hbar\hat{\Omega}^{(nl)} & \equiv -\int d\vR \ \vE^{(l)}(\vR,t)\cdot \hP^{(n)}(\vR)  \\
			&= -2\mu_0 \mu_{ge}^2\hsigma_x^{(n)} \real{\omega^2\rho_{ge}^{(l)}(t) G_{nl} }_{\omega = \omega_{0}} \theta\left(t - \frac{R_{nl}}{c}\right)\\
			&= -2\mu_0 \mu_{ge}^2 \omega_0^2\hsigma_x^{(n)} \Big\{ \real{\rho_{ge}^{(l)}(t)}\real{ G_{nl} } - \imag{\rho_{ge}^{(l)}(t)}\imag{ G_{nl} } \Big \}_{\omega = \omega_{0}}\theta\left(t - \frac{R_{nl}}{c}\right)
			\label{eq:Omega_nl_3}
			\end{align}
		\end{subequations}
		\end{widetext}
		where  $G_{nl} = G_{ln}$ is defined in Eq. \eqref{eq:Greens-short-hand}, $\real{ G_{nl} }$ and $\imag{ G_{nl} }$ are defined in Eq. \eqref{eq:greens-functions-expression-2}. $\theta\left(t - \frac{R_{nl}}{c}\right)$ denotes the Heaviside step function which is required to preserve causality.

		Finally, using Eqs. \eqref{eq:Eperp-expand} and \eqref{eq:intPP-point-dipoles}, the transverse interaction between sites $n$ and $l$ reads
				\begin{subequations} 
					\label{eq:Omega_nl_perp}
			\begin{align}
			\hbar\hat{\Omega}_{\perp}^{(nl)} & \equiv -\int d\vR \ \vE_{\perp}^{(l)}(\vR,t)\cdot \hP^{(n)}(\vR)  \\
			&= \hbar\hat{\Omega}^{(nl)} - \hat{v}^{(nl)}_{\text{Coul}}(t) 
			\label{eq:Omega_nl_perp_2}
			\end{align}
		\end{subequations}
	    For the case of $n = l$, we apply Eqs. \eqref{eq:real-imag-parts}, \eqref{eq:Omega_nl_3}, and \eqref{eq:intPP-point-dipoles}, noting that the terms involving $\real{\rho_{ge}(t)}$ cancel. Then Eq. \eqref{eq:Omega_nl_perp_2} becomes
		\begin{equation}
			\label{eq:Omega_nn_perp}
		\begin{aligned}
		\hbar\hat{\Omega}_{\perp}^{(nn)} = - \hbar\kFGR\imag{\rho_{ge}^{(n)}(t)}\hsigma_x^{(n)}
		\end{aligned}
		\end{equation}		
		where $\kFGR$ is defined in Eq. \eqref{eq:kFGR}.
		
		Thus, in the  end, provided we can make the assumption in Eq. \eqref{eq:rho_ge_retarded}, we have obtained an analytical and EM-free form of Hamiltonian \#I:
		\begin{equation}
		\begin{aligned}
		\hH_{sc}^{I} = \sum_{n = 1}^{N}\hH_s^{(n)}  + \hbar\hat{\Omega}_{\text{in}}^{(n)} + \sum_{n < l} \hV_{\text{Coul}}^{(nl)} + \sum_{nl} \hbar\hat{\Omega}_{\perp}^{(nl)}
		\end{aligned}
		\end{equation}	
		where the analytical expressions of $\hV_{\text{Coul}}^{(nl)} $, $\hbar\hat{\Omega}_{\perp}^{(nl)}$ and $\hbar\hat{\Omega}_{\perp}^{(nn)}$ are defined in Eqs. \eqref{eq:Vdd-point-dipoles},  \eqref{eq:Omega_nl_perp} and \eqref{eq:Omega_nn_perp}; $ \hbar\hat{\Omega}_{\text{in}}^{(n)}$ denotes the coupling between molecule $n$ with the incoming field.

		\subsection{Analytical and EM-Free Form of Hamiltonian \#II}
		By following the procedure above, we can also obtain an EM-free form for Hamiltonian \#II:
		\begin{equation}
		\label{eq:H-II-point-dipole}
		\begin{aligned}
		\hH_{sc}^{II} =  \sum_{n = 1}^{N} \hH_s^{(n)}  + \hbar\hat{\Omega}_{\text{in}}^{(n)} + \hbar\hat{\Omega}_{\perp}^{(nn)} +   \sum_{n\neq l} \hbar\hat{\Omega}^{(nl)}
		\end{aligned}
		\end{equation}
		where $\hbar\hat{\Omega}^{(nl)}$ and $\hbar\hat{\Omega}_{\perp}^{(nn)}$  are defined in Eqs. \eqref{eq:Omega_nl} and \eqref{eq:Omega_nn_perp}. The analytical and EM-free forms of Hamiltonians \#I and \#II allow us to perform simulations of coupled light-matter interactions with negligible computational cost; the propagation of the EM fields on a grid is no longer necessary.
		
		\subsection{Hamiltonians for a Pair of Two-Level Systems}
		In this manuscript, we have presented results for a minimalistic many-site model --- a pair of identical TLSs (labeled as D and A). For convenience, we now report the analytical and EM-free form of Hamiltonian \#I for the pair of TLSs.
		
		Let us define $\hbar\Omega^{(n)}_{\text{in}}$ and $\hbar\Omega^{(nl)}_{\perp}$ as the norms of the corresponding operators that have already been defined, i.e., $\hbar\hat{\Omega}^{(n)}_{\text{in}} = \hbar\Omega^{(n)}_{\text{in}}\hsigma_x^{(n)}$, 
		$\hbar\hat{\Omega}^{(nl)}_{\perp} = \hbar\Omega^{(nl)}_{\perp}\hsigma_x^{(n)}$ (for $n, l = D, A$), where $\hbar\hat{\Omega}^{(nl)}_{\perp}$ is defined in Eqs. \eqref{eq:Omega_nl_perp} and \eqref{eq:Omega_nn_perp}. Then, for the pair of TLSs, 
		\begin{equation}
		    \hH_{sc}^{I}=
		    \begin{pmatrix}
		    0 & V_A & V_D & v_{dd}\\
		    V_A & \hbar\omega_0 & v_{dd} & V_D \\
		    V_D & v_{dd} & \hbar\omega_0 & V_A \\
		    v_{dd} & V_D & V_A & 2\hbar\omega_0
		    \end{pmatrix}
		\end{equation}
		Here, $v_{dd} = \frac{ \eta_{3}^{(DA)}}{4\pi \epsilon_0 R_{DA}^3}$, $V_D = \hbar\Omega^{(D)}_{\text{in}} + \hbar\Omega^{(DD)}_{\perp} + \hbar\Omega^{(DA)}_{\perp}$,  $V_A = \hbar\Omega^{(A)}_{\text{in}} + \hbar\Omega^{(AA)}_{\perp} + \hbar\Omega^{(AD)}_{\perp}$.
		
		Similarly, Hamiltonian \#II for  a pair of TLSs reads
		\begin{equation}
		    \hH_{sc}^{II}=
		    \begin{pmatrix}
		    0 & V_A' & V_D' & 0\\
		    V_A' & \hbar\omega_0 & 0 & V_D' \\
		    V_D' & 0 & \hbar\omega_0 & V_A' \\
		    0 & V_D' & V_A' & 2\hbar\omega_0
		    \end{pmatrix}
		\end{equation}	
		where $V_D' = \hbar\Omega^{(D)}_{\text{in}} + \hbar\Omega^{(DD)}_{\perp} + \hbar\Omega^{(DA)}$, and $V_A' = \hbar\Omega^{(A)}_{\text{in}} + \hbar\Omega^{(AA)}_{\perp} + \hbar\Omega^{(AD)}$. Here, as above, we have defined $\hbar\Omega^{(nl)}$ as the norm of $\hbar\hat{\Omega}^{(nl)}$ [defined in Eq. \eqref{eq:Omega_nl}], i.e., $\hbar\hat{\Omega}^{(nl)} = \hbar\Omega^{(nl)}\hsigma_x^{(n)}$ (for $n \neq l$).
		
		Finally, for  the hybrid Hamiltonian  for  a pair of TLSs [see Eq. \eqref{eq:H-hybrid}], the Hamiltonian reads:
		\begin{equation}\label{eq:Hhyb_2TLS}
	\hH_{sc}^{hyb}=
\begin{pmatrix}
0 & V_A & V_D & 0\\
V_A & \hbar\omega_0 & v_{dd} & V_D' \\
V_D & v_{dd} & \hbar\omega_0 & V_A' \\
0 & V_D' & V_A' & 2\hbar\omega_0
\end{pmatrix}
\end{equation}

		\section{The Lehmberg-Agarwal master equation (LAME)}
		\label{Appendix: LAME}
	
		\setcounter{equation}{0}
		\setcounter{figure}{0}
		\setcounter{table}{0}
		\renewcommand{\theequation}{B\arabic{equation}}
		\renewcommand{\thefigure}{B\arabic{figure}}
		 For $N$ identical TLSs, the Lehmberg-Agarwal master equation\cite{Lehmberg1970,Agarwal1974} (LAME) is the standard theory to describe the  reduced dynamics of the electronic degrees of freedom in an open quantum environment. Formally, one can derive the LAME
		 by taking the Born-Markov approximation from QED and a rotating wave approximation (RWA), leading to
		 \begin{widetext}
		\begin{equation}\label{eq:Linblad}
		\begin{aligned}
				\frac{d}{dt}\hrho_N(t) = &-\frac{i}{\hbar}\left[\sum_{n = 1}^{N}\hH_s^{(n)} + \hbar\hat{\Omega}^{(n)}_{\text{in}}, \hrho_N\right]  
		- i\sum_{n \neq l}^{N} b_{nl}\left[\hsigma_{+}^{(n)}\hsigma_{-}^{(l)}, \hrho_N\right]
		+ \Lrhosub{L}{N}
		\end{aligned}
		\end{equation}
		\end{widetext}
		where the dissipative term $\Lrhosub{L}{N}$ is called the Lindbladian:
		\begin{widetext}
		\begin{equation}\label{eq:Lindbladian}
		\begin{aligned}
					\Lrhosub{L}{N}
		= \sum_{nl} a_{nl}\Big\{  & \hsigma_{-}^{(l)}\hrho_N \hsigma_{+}^{(n)} 
		- \frac{1}{2}\hsigma_{+}^{(n)} \hsigma_{-}^{(l)} \hrho_N 
	- \frac{1}{2}\hrho_N \hsigma_{+}^{(n)} \hsigma_{-}^{(l)} 
		\Big\}
		\end{aligned}
		\end{equation}
		\end{widetext}
		Here, $\hrho_N$ denotes the $N$-body density operator, 
		$\hbar\hat{\Omega}^{(n)}_{\text{in}}$ denotes the coupling between the incoming E-field and molecule $n$, and the $a_{nl}$ and $b_{nl}$ terms describe the collective damping and the collective level shifts,
		which are defined as
		\begin{widetext}
		\begin{subequations}\label{eq:a_and_b_nl}
			\begin{align}
			a_{nl} &=
			\begin{cases}
			\kFGR^{(n)} = \frac{\omega_0^3|\mu_{ge}^{(n)}|^2}{3\pi \hbar c^3\epsilon_0} ,& \text{if } l = n\\
			\frac{\omega_0^3\mu_{ge}^{(n)}\mu_{ge}^{(l)}}{2\pi \hbar c^3\epsilon_0}\left[
			\frac{\sin x_{nl}}{x_{nl}}\eta_1^{(nl)}
			+ \frac{\cos x_{nl}}{x_{nl}^2}\eta_3^{(nl)}
			- \frac{\sin x_{nl}}{x_{nl}^3}\eta_3^{(nl)}
			\right],              & \text{otherwise}
			\end{cases}
			\label{eq:a_nl}
			\\
			b_{nl} &=  \frac{\omega_0^3\mu_{ge}^{(n)}\mu_{ge}^{(l)}}{4\pi \hbar c^3\epsilon_0}\left[
			-\frac{\cos x_{nl}}{x_{nl}}\eta_1^{(nl)}
			+ \frac{\sin x_{nl}}{x_{nl}^2}\eta_3^{(nl)}
			+ \frac{\cos x_{nl}}{x_{nl}^3}\eta_3^{(nl)}
			\right]\left(1 - \delta_{nl}\right)
			\label{eq:b_nl}
			\end{align}
		\end{subequations}
	\end{widetext}
		where the dimensionless intermolecular separation $x_{nl}$ is defined as $ x_{nl} \equiv \omega_0 R_{nl}/c$, and the Kronecker delta function $\delta_{nl}$ equals to $1$ if $n = l$ and equals $0$ otherwise.
		As might be guessed from the structures of $a_{nl}$ and $b_{nl}$ (that contain both $1/R^6$ and $1/R^2$ terms as well as $\kFGR$), LAME can accurately capture the time-resolved RET dynamics between a pair of TLSs at both short and long range.

		Although not the focus of this paper, when modeling dynamics with LAME, one key problem is that for a system with $N$ TLSs, the method requires one to build an exponentially large many-body density matrix of size $2^N$ during the course of a simulation and update the Lindbladian at every time step. As a result, LAME is usually applied only to a few TLSs.
		Furthermore, applying LAME for inhomogeneous systems (not in vacuum) is not obvious; and more generally, like any master equation, LAME is accurate only in the limit of weak light-matter coupling.
		
		\subsection{Connecting Ehrenfest Dynamics with LAME}	
		Above, in Fig. \ref{fig:RET_dynamics}, we have observed that LAME agrees with Ehrenfest \#II if the TLSs are weakly excited. Let us now analytically show that LAME can indeed be connected to Ehrenfest \#II after some approximations are made. We will start from the EM-free form of Hamiltonian \#II [see Eq. \eqref{eq:H-II-point-dipole}] and take the RWA form of Hamiltonian \#II.

		We assume that $\text{Im}\left[\rho_{ge}\right] \approx \widetilde{\rho}_{ge} \text{Im}\left[e^{i\omega_0 t} \right]$, 
		where $\widetilde{\rho}_{ge} \equiv \rho_{ge}e^{-i\omega_0 t}$ is a slowly varying variable compared with the time scale of $\omega_0^{-1}$. With this assumption, 
		we obtain from Eq. \eqref{eq:Omega_nn_perp} the  RWA form of $\hbar\hat{\Omega}_{\perp}^{(nn)}$:
		\begin{equation}\label{eq:Omega_nn_RWA}
		\begin{aligned}
		\hat{\Omega}^{(nn)}_{\perp, \text{RWA}} &= \frac{i}{2}\kFGR^{(n)} \widetilde{\rho}_{ge}^{(n)}\left[ e^{i\omega_0t}\hsigma_{-}^{(n)}
		-e^{-i\omega_0t}\hsigma_{+}^{(n)}
		\right] 
		\end{aligned}
		\end{equation}
		Similarly, for $\hbar\hat{\Omega}^{(nl)} $ ($l\neq n$) in Eq. \eqref{eq:Omega_nl_3}, the corresponding RWA form reads
		\begin{equation}\label{eq:Omega_nl_RWA}
		\begin{aligned}
		\hat{\Omega}^{(nl)}_{\text{RWA}} = \ &-\frac{1}{2}\frac{\omega_0^3 \mu_{ge}^{(n)} \mu_{ge}^{(l)}}{2 \pi \hbar c^3\epsilon_0}\left[\frac{\eta_1}{x_{nl}} - \frac{i\eta_3}{x_{nl}^2} - \frac{\eta_3}{x_{nl}^3}\right]\widetilde{\rho}_{ge}^{(l)} \\
		&\times e^{i\omega_0\left(t - \frac{r_{nl}}{c}\right)}\theta\left(t-\frac{r_{nl}}{c}\right)\hsigma_{-}^{(n)} + \text{c.c.} 
		\end{aligned}
		\end{equation}
		where c.c. denotes the complex conjugate.
		Let us make the following definitions
		\begin{widetext}
		\begin{equation}\label{eq:gamma_nl}
		\begin{aligned}
		\gamma_{nn} &= i\kFGR^{(n)} = i\frac{\omega_0^3|\mu_{ge}^{(n)}|^2}{3\pi \hbar c^3\epsilon_0} \\
		\gamma_{nl} &= -\frac{\omega_0^3 \mu_{ge}^{(n)} \mu_{ge}^{(l)}}{2\pi \hbar c^3\epsilon_0} \left[
		\frac{e^{ - ix_{nl}}}{x_{nl}}\eta_1
		- i\frac{e^{-ix_{nl}}}{x_{nl}^2} \eta_3
		- \frac{e^{- ix_{nl}}}{x_{nl}^3}\eta_3
		\right]   \theta\left(t-\frac{r_{nl}}{c}\right),
		\end{aligned}
		\end{equation}
		\end{widetext}
		so that all light-matter couplings can be rewritten in a uniform way (for both $l = n$ and $l \neq n$)
		\begin{equation}\label{eq:Omega_nl_RWA_final}
		\hat{\Omega}_{\text{RWA}}^{(nl)} = \frac{1}{2}
		\left[
		\gamma_{nl}
		\tr{\hrho^{(l)}\hsigma_{+}^{(l)}}\hsigma_{-}^{(n)} +
		\gamma_{nl}^{\ast}
		\tr{\hrho^{(l)}\hsigma_{-}^{(l)}}\hsigma_{+}^{(n)} 
		\right]
		\end{equation}
		Because $\gamma_{nl}$ in Eq. \eqref{eq:gamma_nl} is a c-number, the real and imaginary parts of $\gamma_{nl}$ contribute differently to the electronic dynamics, i.e., the imaginary part of $\gamma_{nl}$ leads to dissipation while the real part should be a level shift. Thus, it is necessary to separate the real and imaginary parts of $\gamma_{nl}$:
		\begin{equation}\label{eq:a_b_Eh}
		\begin{aligned}
		a_{nl}' &= \text{Im}\left[\gamma_{nl}\right]  \\
		b_{nl}' &= \frac{1}{2}\text{Re}\left[\gamma_{nl}\right] .
		\end{aligned}
		\end{equation}
		With these definitions, 
		we can further rewrite Eq. \eqref{eq:Omega_nl_RWA_final}  as
		\begin{equation}
		\begin{aligned}
		\hat{\Omega}_{\text{RWA}}^{(nl)} &= b_{nl}'\left[\tr{\hrho^{(l)}\hsigma_{+}^{(l)}}\hsigma_{-}^{(n)} + \tr{\hrho^{(l)}\hsigma_{-}^{(l)}}\hsigma_{+}^{(n)}\right] \\
		& +\frac{i}{2} a_{nl}' \left[\tr{\hrho^{(l)}\hsigma_{+}^{(l)}}\hsigma_{-}^{(n)} - \tr{\hrho^{(l)}\hsigma_{-}^{(l)}}\hsigma_{+}^{(n)}\right] 
		\end{aligned}
		\end{equation}
		and the RWA form of Hamiltonian \#II becomes	
		\begin{equation}
		\hH_{\text{II, RWA}}^{(n)} = \hH_s^{(n)} + \hbar\hat{\Omega}_{\text{in}}^{(n)} + \sum_{l=1}^{N} \hbar\hat{\Omega}_{\text{RWA}}^{(nl)}
		\end{equation}
		The expressions for the $a_{nl}'$ and $b_{nl}'$ coefficients reported here [Eq. \eqref{eq:a_b_Eh}] are exactly the same as the coefficients of LAME [$a_{nl}$ and $b_{nl}$ in Eq. \eqref{eq:a_and_b_nl}], provided that the step function $\theta\left(t-\frac{r_{nl}}{c}\right)$ in Eq. \eqref{eq:gamma_nl} (or causality) is ignored. The exact agreement between these distinct coefficients clearly suggests a consistency between Hamiltonian \#II and LAME. In fact,
		one can show that,  if the step function (or causality) is ignored, $\hH_{\text{II, RWA}}^{(n)}$ is exactly the effective mean-field Hamiltonian $\hH_{\text{mf}}^{(n)}$ of MF-LAME defined by
		\begin{equation}
		\label{eq:drhodt_Mean}
		\frac{d}{dt}\hrho_N(t) = -\frac{i}{\hbar}\left[\sum_{n = 1}^{N}\hH_{\text{mf}}^{(n)}, \  \hrho_N\right]
		\end{equation}
		Eq. \eqref{eq:drhodt_Mean} can be derived from Eq. \eqref{eq:Linblad} by supposing $\hrho_N(t) = \hrho^{(1)}(t) \otimes \cdots \otimes \hrho^{(N)}(t)$ is valid at any time $t$ and then tracing out $N-1$ degrees of freedom to form the one-body reduced density operator; see Ref. \citenum{Breuer2007} for a detailed procedure.

		\section{The RET Rate}
		\label{Appendix:RET_rate}

\setcounter{equation}{0}
\setcounter{figure}{0}
\setcounter{table}{0}
\renewcommand{\theequation}{C\arabic{equation}}
\renewcommand{\thefigure}{C\arabic{figure}}

In Fig. \ref{fig:RET_rate}, we have compared the RET rate calculated by different approaches. For the sake of completeness, we will now briefly review the RET rate theory. Furthermore,  we will also analytically calculate the short-time result of $\rho_{ee}^{(A)}(t)$ as propagated with Ehrenfest dynamics by Hamiltonian \#II, confirming the numerical calculations in Fig. \ref{fig:RET_rate} and the finite-difference time-domain (FDTD) simulation results in Ref. \citenum{Li2018Tradeoff}.

\subsection{Perturbative QED Result}
According to the standard perturbative QED calculations\cite{Salam2012,Salam2018},
in the weak coupling limit,
the RET rate between a pair of TLSs [donor (D) + acceptor (A)] can be calculated by Fermi's golden rule:
\begin{equation}\label{eq:kET}
\kET = \frac{2\pi}{\hbar}  \left| M(\omega_0, R_{DA}) \right|^2 \rho_f
\end{equation}
Here, $\rho_f$ denotes the density of states for the final state,
$M(\omega, R_{DA})$ denotes the transition matrix element between the final state and initial state, and $R_{DA}$ denotes the separation between the donor and acceptor. 
In order to evaluate $M(\omega, R_{DA})$, let us use the notation $\ket{nmk}$ to represent the donor in state $\ket{n}$, the acceptor in state $\ket{m}$, and the photon in state $\ket{k}$. Whereas the initial state $\ket{eg0}$ and the final state $\ket{ge0}$ do not  couple directly through the EM field, one can show that they are coupled at second order. To do so, one simply expands the initial state to first order, i.e., $\ket{eg0}\rightarrow \ket{\psi_{eg0}} =  \ket{eg0} +\sum\limits_k \frac{\ket{ggk}\avg{ggk| \hH_{\text{int}} | eg0}}{\omega_k - \omega_0}$, where $\hH_{\text{int}}$ denotes the interaction Hamiltonian. Then, to second order in the interaction, the key contribution should come from the state $\ket{ggk}$ where the photon has frequency  $\omega_k = \omega_0$, i.e., the photon energy should equal the energy gap for both the donor and acceptor. To second-order in the perturbation, the coupling matrix
$M(\omega, R_{DA})$ takes the form (after a few integrations in 3D)\cite{AkbarSalam2010}:
\begin{widetext}
\begin{equation}
M(\omega, R_{DA}) = \frac{\omega^3 \mu_{ge}^{(D)} \mu_{ge}^{(A)}}{4\pi c^3\epsilon_0} 
\left[
-\frac{c\eta_1^{(DA)}}{\omega R_{DA}}
-i\frac{c^2 \eta_3^{(DA)}}{\omega^2 R^2_{DA}}
+ \frac{c^3\eta_3^{(DA)}}{\omega R^3_{DA}}
\right]e^{i\frac{\omega R_{DA}}{c}}
\end{equation} 
\end{widetext}

At this point, consider the density of states for the acceptor ($\rho_f$ in Eq. \eqref{eq:kET}). If there are no vibrations (or other electronic degrees of freedoms), then over the time scale $\omega_0^{-1} \ll t \ll \kFGR^{-1}$, [where $\kFGR$ denotes the spontaneous emission (SE) rate for a single TLS] 
there can be no true rate of energy transfer. Instead, one will find large oscillations back and forth. At very short times, the excited state population for the acceptor  is simply
\begin{equation}\label{eq:rho22_A_QED}
\begin{aligned}
\rho_{ee, \text{QED}}^{(A)}(t) =& \frac{\rho_{ee}^{(D)}(0)}{\hbar^2}|M(\omega_0, R_{DA})|^2 \left(t - \frac{R_{DA}}{c}\right)^2 \\
& \times \theta\left(t - \frac{R_{DA}}{c}\right)
\end{aligned}
\end{equation}
where $\rho_{ee}^{D}(0)$ denotes the initial excited state population for the donor.

\subsection{Analytical RET Rate by Hamiltonian \#II}
According to  Ehrenfest dynamics with Hamiltonian \#II,
the equations of motion for the acceptor (A) read 
\begin{subequations}\label{eq:eq_of_motion_acceptor}
	\begin{align}
	\frac{d\rho_{ee}^{(A)}}{dt} &= -\frac{2}{\hbar}\vmu\cdot \vE\text{Im}\left[\rho_{ge}^{(A)}\right] \label{eq:dP2dt_acceptor} \\
	\frac{d\rho_{ge}^{(A)}}{dt} &= i\omega_0\rho_{ge}^{(A)}+ \frac{i}{\hbar}\vmu\cdot \vE\left(\rho_{ee}^{(A)} - \rho_{gg}^{(A)}\right)\label{eq:drho12dt_acceptor}
	\end{align}
\end{subequations}
where $\vmu = \mu_{ge} \ve_d$.
At short times, 
because the excited state population for the acceptor is much smaller than the donor, the
EM fields that are felt by the acceptor predominately come from the donor. Thus, at short times, we can neglect the donor's population decay, [i.e., $\rho_{ge}^{(D)}(t) \approx \rho_{ge}^{(D)}(0)e^{i\omega_0 t}$], so that the light-matter coupling term for the acceptor is just $-\vmu\cdot\vE = \hbar\Omega \approx \hbar\Omega^{(AD)}$, where $\hbar\Omega^{(AD)}$ is defined by $\hbar\hat{\Omega}^{(AD)} = \hbar\Omega^{(AD)}\hsigma_x^{(A)}$ [see Eq. \eqref{eq:Omega_nl_RWA}]. Furthermore, according to the RWA in Eq. \eqref{eq:Omega_nl_RWA},
\begin{equation}\label{eq:Omega_donor_RWA}
\begin{aligned}
\Omega_{\text{RWA}}^{(AD)} = \ &-\frac{\mu_{ge}^2\omega_0^3}{4\pi\hbar\epsilon_0 c^3}|\widetilde{\rho}_{ge}^{(D)}(0)|\left[\frac{\eta_1}{x} - i\frac{\eta_3}{x^2} - \frac{\eta_3}{x^3}\right] \\
&\times e^{i(\omega_0 t - x)}\theta\left(t-\frac{R_{DA}}{c}\right)
\end{aligned}
\end{equation}
Because there is no non-Hamiltonian dissipative term in Ehrenfest dynamics, purity is strictly conserved, i.e.,
\begin{equation}\label{eq:rho12_acceptor}
\rho_{ge}^{(A)} = \sqrt{\rho_{ee}^{(A)}\rho_{gg}^{(A)}}e^{i(\omega_0t + \varphi)}
\end{equation}
where $\sqrt{\rho_{ee}^{(A)}\rho_{gg}^{(A)}}$ is slowing varying compared with the time scale of $2\pi/\omega_0$, and $\varphi$ is the initial phase for the acceptor.
By further substituting Eqs. \eqref{eq:Omega_donor_RWA} and \eqref{eq:rho12_acceptor} into Eq. \eqref{eq:drho12dt_acceptor}, we obtain
\begin{equation}\label{eq:dP2dt_Acceptor}
\begin{aligned}
\frac{d}{dt}\sqrt{\rho_{ee}^{(A)}\rho_{gg}^{(A)}} = & \ i\frac{\mu_{ge}^2\omega_0^3}{4\pi \epsilon_0 c^3}\rho_{ge}^{(D)}(0)\left[\frac{\eta_1}{x} - i\frac{\eta_3}{x^2} - \frac{\eta_3}{x^3}\right] \\
&\times e^{i(x+\varphi)}\theta\left(t-\frac{R_{DA}}{c}\right)
\end{aligned}
\end{equation}
For short times, the acceptor is  not strongly excited, i.e., $\sqrt{\rho_{ee}^{(A)}\rho_{gg}^{(A)}} \approx \sqrt{\rho_{ee}^{(A)}}$, so that Eq. \eqref{eq:dP2dt_Acceptor} is easily integrated [with $\int_0^{t} \theta(t'-T) dt' = (t-T)\theta(t-T)$]:
\begin{equation}\label{eq:rho22_short_time_Eh_Appendix}
\begin{aligned}
\rho_{ee, \text{Eh}}^{(A)}(t) &= \rho_{gg}^{(D)}(0)\rho_{ee}^{(D)}(0)\left|\frac{\omega_0^3\mu_{ge}^2}{4\pi\epsilon_0c^3} \left[\frac{\eta_1}{x} - i\frac{\eta_3}{x^2} - \frac{\eta_3}{x^3}\right]\right|^2 \\
&\times \left(t - \frac{R_{DA}}{c}\right)^2\theta\left(t-\frac{R_{DA}}{c}\right) 
\\
& = \rho_{gg}^{(D)}(0) \rho_{22, \text{QED}}^{(A)}(t)
\end{aligned}
\end{equation}
In other words, Ehrenfest predicts that the excited state population on the acceptor will be just $\rho_{gg}^{(D)}(0)$ times the perturbative QED result [$ \rho_{ee, \text{QED}}^{(A)}(t)$ in Eq. \eqref{eq:rho22_A_QED}]. When the donor is near the ground state, i.e., $\rho_{gg}^{(D)}(0)\rightarrow 1$, Ehrenfest dynamics [with Hamiltonian \#II] exactly recovers the perturbative QED result. Note that the analytical derivations here exactly agree with our previous FDTD simulations\cite{Li2018Tradeoff}.

		\end{appendices}

	
	%

\end{document}